\documentclass[toc]{PoS}

\usepackage{graphicx}

\PoS{PoS(LAT2005)012}

\def\OMIT#1{{}}
\def\lsim{\mathrel{\!\mathpalette\vereq<}\!}
\def\gsim{\mathrel{\!\mathpalette\vereq>}\!}
\def\vereq#1#2{\lower3.5pt\vbox{\baselineskip1.5pt \lineskip1.5pt
\ialign{$#1\hfill##\hfil$\crcr#2\crcr\sim\crcr}}}

\def\lqcd{\Lambda_{\rm QCD}}

\def\d{{\rm d}}
\def\ov{\overline}

\def\TeV{{\rm TeV}}
\def\GeV{{\rm GeV}}
\def\MeV{{\rm MeV}}
\def\rhobar{\bar\rho}
\def\etabar{\bar\eta}

\newcommand{\Bbar}{\,\overline{\!B}}
\newcommand{\Dbar}{\,\overline{\!D}}
\newcommand{\Kbar}{\,\overline{\!K}}
\def\B0bar{\Bbar{}^0}
\def\D0bar{\Dbar{}^0}
\def\K0bar{\Kbar{}^0}
\def\nslash{n\hspace{-5pt}\slash}
\def\nbarslash{\bar n\hspace{-5pt}\slash}
\def\vslash{v\!\!\!\!\slash}

\newcommand{\nn}{\nonumber}
\newcommand{\beq}{\begin{equation}}
\newcommand{\eeq}{\end{equation}}
\newcommand{\beqa}{\begin{eqnarray}}
\newcommand{\eeqa}{\end{eqnarray}}

\title{The CKM matrix and CP violation (in the continuum approximation)}

\ShortTitle{The CKM matrix and CP violation}

\author{\speaker{Zoltan Ligeti}\\
Ernest Orlando Lawrence Berkeley National Laboratory,
University of California, Berkeley, CA 94720\hspace*{-1cm}\\
and\\
Center for Theoretical Physics, Massachusetts Institute of Technology,
Cambridge, MA 02139\\
E-mail: \email{ligeti@lbl.gov}}

\abstract{The first part of this talk reviews recent developments in flavor
physics that can be made without detailed understanding of hadronic physics,
driven by the data.  The error of $\sin2\beta$ has shrunk below 5\%, and the
measurements of $\alpha$ and $\gamma$ have reached interesting precisions.  For
the first time, there are significant constraints on the deviations from the
standard model in $B-\Bbar$ mixing and in $b\to s$ and $b\to d$ transitions.  In
the second part, I review some theoretical developments for exclusive
semileptonic and nonleptonic $B$ decays that have become possible using the
soft-collinear effective theory.  I concentrate on topics where the recent
progress has model independent implications for interpreting the data. \hfill
\mbox{\footnotesize LBNL-59133, MIT-CTP 3715}}

\FullConference{XXIIIrd International Symposium on Lattice Field Theory\\
25-30 July 2005\\
Trinity College, Dublin, Ireland}

\begin{document}

\section{Introduction}

In the last few years the study of $CP$ violation and flavor physics has
undergone dramatic developments.  While for 35 years, until 1999, the only
unambiguous measurement of $CP$ violation (CPV) was $\epsilon_K$~\cite{Kcpv},
the constraints on the Cabibbo-Kobayashi-Maskawa (CKM) matrix~\cite{C,KM}
improved tremendously since the $B$ factories turned on.  The error of
$\sin2\beta$ is now below 5\%, and a new set of measurements started to give the
best constraints on the CKM parameters.

In the standard model (SM), the masses and mixings of quarks originate from
their Yukawa interactions with the Higgs condensate.  We do not understand the
hierarchy of the quark masses and mixing angles.  Moreover, if there is new
physics (NP) at the TeV scale, as suggested by the hierarchy problem, then it is
not clear why it has not shown up in flavor physics experiments. A four-quark
operator $(s\bar d)^2/\Lambda_{\rm NP}^2$ with ${\cal O}(1)$ coefficient would
give a contribution exceeding the measured value of $\epsilon_K$ unless
$\Lambda_{\rm NP} \gsim 10^4\,\TeV$.  Similarly, $(d\bar b)^2/\Lambda_{\rm
NP}^2$ yields $\Delta m_{B_d}$ above its measured value unless $\Lambda_{\rm NP}
\gsim 10^3\,\TeV$.  Flavor physics provides significant constraints on NP model
building; for example generic SUSY models have 43 new $CP$ violating
phases~\cite{Haber:1997if,Nir:2001ge}, and we already know that many of them
have to be suppressed not to contradict the experimental data.

Flavor and $CP$ violation were excellent probes of new physics in the past: (i)
the absence of $K_L\to \mu^+\mu^-$ predicted the charm quark; (ii) $\epsilon_K$
predicted the third generation; (iii) $\Delta m_K$ predicted the charm mass;
(iv) $\Delta m_B$ predicted the heavy top mass.  From these measurements we knew
already before the $B$ factories turned on that if there is NP at the TeV scale,
it must have a very special flavor and $CP$ structure to satisfy these
constraints.  So what does the new data tell us?  

Sections 2--4 summarize the status of $CP$ violation measurements and their
implications within and beyond the SM, concentrating on measurements where the
data can be interpreted without detailed understanding of the hadronic physics. 
Sections 5--7 deal with some recent model independent theoretical developments
and their implications.

\subsection{Testing the flavor sector}

The only interaction that distinguishes between the fermion generations is their
Yukawa couplings to the Higgs condensate.  This sector of the SM contains 10
physical quark flavor parameters, the 6 quark masses and the 4 parameters in the
CKM matrix: 3 mixing angles and 1 $CP$ violating phase (for reviews, see,
e.g.,~\cite{Nir:2001ge,Ligeti:2003fi}).  Therefore, the SM predicts intricate
correlations between dozens of different decays of $s$, $c$, $b$, and $t$
quarks, and in particular between $CP$ violating observables.  Possible
deviations from CKM paradigm may upset some predictions:
\begin{itemize}\vspace*{-8pt}\itemsep -2pt
\item Subtle (or not so subtle) changes in correlations, e.g., constraints from
$B$ and $K$ decays inconsistent, or $CP$ asymmetries not equal in $B\to \psi
K_S$ and $B\to \phi K_S$, etc.;
\item Flavor-changing neutral currents at an unexpected level, e.g., $B_s$
mixing incompatible with SM,  enhanced $B_{(s)}\to \ell^+ \ell^-$, etc.;
\item Enhanced (or suppressed) $CP$ violation, e.g., in $B\to K^*\gamma$ or
$B_s\to \psi \phi$.
\end{itemize}\vspace*{-10pt}

\begin{figure}[t]
\centerline{\includegraphics*[width=.4\textwidth]{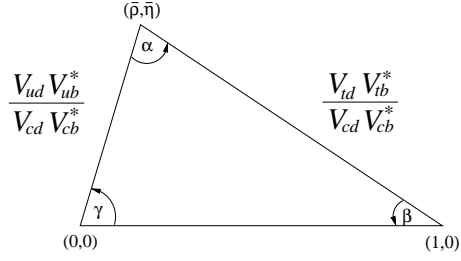}}
\caption{Sketch of the unitarity triangle.}
\label{fig:triangle}
\end{figure}

The goal of the program is not just to determine SM parameters as precisely as
possible, but to test by many overconstraining measurements whether all
observable flavor-changing interactions can be explained by the SM, i.e., by
integrating out virtual $W$ and $Z$ bosons and quarks.  It is convenient to use
the Wolfenstein parameterization\footnote{We use the following
definitions~\cite{Wolfenstein:1983yz,Buras:1994ec,Charles:2004jd}, so that the
apex of the unitarity triangle in Fig.~\ref{fig:triangle} is exactly
$\rhobar,\etabar$:
\beq\label{laredef}
\lambda = \frac{|V_{us}|}{\sqrt{|V_{ud}|^2+|V_{us}|^2}} \,, \qquad
A = \frac1\lambda \bigg|{V_{cb}\over V_{us}}\bigg| \,, \qquad
V_{ub}^* = A\lambda^3 (\rho + i\eta)
  = {A\lambda^3(\rhobar+i\etabar)\sqrt{1-A^2\lambda^4}\over \sqrt{1-\lambda^2}
  [1-A^2\lambda^4(\rhobar+i\etabar)]} \,.
\eeq} of the CKM matrix,
\beq\label{ckmdef}
V_{\rm CKM} = \pmatrix{ V_{ud} & V_{us} & V_{ub} \cr
  V_{cd} & V_{cs} & V_{cb} \cr
  V_{td} & V_{ts} & V_{tb} } =
\pmatrix{ 1-\frac{1}{2}\lambda^2 & \lambda & A\lambda^3(\rhobar-i\etabar) \cr
  -\lambda & \!\!1-\frac{1}{2}\lambda^2\!\! & A\lambda^2 \cr
  A\lambda^3(1-\rhobar-i\etabar) & -A\lambda^2 & 1} + \ldots \,,
\eeq
which exhibits its hierarchical structure by expanding in $\lambda \simeq 0.23$,
and is valid to order $\lambda^4$.  The unitarity of the CKM matrix implies
$\sum_i V_{ij} V_{ik}^* = \delta_{jk}$ and $\sum_j V_{ij} V_{kj}^* =
\delta_{ik}$, and the six vanishing combinations can be represented by triangles
in a complex plane.  The ones obtained by taking scalar products of
neighboring rows or columns are nearly degenerate, so one usually considers
\beq
V_{ud}\, V_{ub}^* + V_{cd}\, V_{cb}^* + V_{td}\, V_{tb}^* = 0 \,.
\eeq
A graphical representation of this is the unitarity triangle, obtained by
rescaling the best-known side to unit length (see Fig.~\ref{fig:triangle}). Its
sides and angles can be determined in many "redundant" ways, by measuring $CP$
violating and conserving observables.  Comparing  constraints on $\rhobar$ and
$\etabar$ provides a  convenient language to compare the overconstraining
measurements.

\subsection{Constraints from $K$ and $D$ decays}
\label{sec:KD}

We knew from the measurement of $\epsilon_K$ that CPV in the $K$ system is at a
level compatible with the SM, as $\epsilon_K$ can be accommodated with an ${\cal
O}(1)$ value of the KM phase~\cite{KM}.  The other observed $CP$ violating
quantity in kaon decay, $\epsilon_K'$, is notoriously hard to interpret, because
for the large top quark mass the electromagnetic and gluonic penguin
contributions tend to cancel~\cite{Flynn:1989iu}, thereby significantly
amplifying the hadronic uncertainties.  At present, we cannot even rule out that
a large part of the measured value of $\epsilon_K'$ is due to NP, and so we
cannot use it to tests the KM mechanism.  In the kaon sector precise tests will
come from the study of $K\to \pi\nu\bar\nu$ decays.  The $K_L\to
\pi^0\nu\bar\nu$ decay is $CP$ violating, and therefore theoretically very
clean, and there is progress in understanding the largest uncertainties in
$K^\pm\to \pi^\pm\nu\bar\nu$ due to charm and light quark
loops~\cite{Isidori:2005xm,Buras:2005gr}.  In this mode three events have been
observed so far, yielding~\cite{Anisimovsky:2004hr}
\beq
{\cal B}(K^+\to \pi^+\nu\bar\nu) = \big(1.5^{+1.3}_{-0.9}\big) \times 10^{-10}\,.
\eeq
This is consistent with the SM within the large uncertainties, but much more
statistics is needed to make definitive tests.  

The $D$ meson system is complementary to $K$ and $B$ mesons, because flavor and
$CP$ violation are suppressed both by the GIM mechanism and by the Cabibbo
angle.  Therefore, CPV in $D$ decays, rare $D$ decays, and $D-\Dbar$ mixing are
predicted to be small in the SM and have not been observed.  This is the only
neutral meson system in which mixing generated by down-type quarks in the SM
(or up-type squarks in SUSY).  The strongest hint for $D^0-\D0bar$ mixing
is the lifetime difference between the $CP$-even and -odd
states~\cite{Yabsley:2003rn}
\beq
y_{CP} = {\Gamma(CP \mbox{ even}) - \Gamma(CP \mbox{ odd}) \over
  \Gamma(CP \mbox{ even}) + \Gamma(CP \mbox{ odd})} = (0.9 \pm 0.4)\%\,.
\eeq
Unfortunately, due to hadronic uncertainties, this central value alone could not
be interpreted as a sign of new physics~\cite{Falk:2001hx}.  At the present
level of sensitivity, CPV or enhanced rare decays would be the only clean signal
of NP in the $D$ sector.

\section{$CP$ violation in $B$ decays and the measurement of $\sin2\beta$}
\label{sec:Bbc}

\subsection{$CP$ violation in decay}
\label{CPVdecay}

This is the simplest form of $CP$ violation, which can be observed in both
charged and neutral meson as well as in baryon decays.  If at least two
amplitudes with nonzero relative weak $(\phi_k)$ and strong $(\delta_k)$ phases
contribute to a decay,
\beq
A_f = \langle f | {\cal H} |B\rangle = 
  \sum_k A_k\, e^{i\delta_k}\, e^{i\phi_k}\,, \qquad
\ov{A}_{\ov f} = \langle \ov f | {\cal H} |\Bbar\rangle = 
  \sum_k A_k\, e^{i\delta_k}\, e^{-i\phi_k}\,,
\eeq
then it is possible that $|{\ov A_{\ov f} / A_f}| \neq 1$, and thus $CP$ is
violated.

This type of $CP$ violation is unambiguously observed in the kaon sector by
$\epsilon'_K \neq 0$, and now it is also established in $B$ 
decays~\cite{Aubert:2004qm,Abe:2005fz},
\beq\label{KpidirectCP}
A_{K^-\pi^+} \equiv {\Gamma(\B0bar\to K^-\pi^+) - \Gamma(B^0\to K^+\pi^-)
  \over \Gamma(\B0bar\to K^-\pi^+) + \Gamma(B^0\to K^+\pi^-)} 
  = -0.115 \pm 0.018\,.
\eeq
This is simply a counting experiment: there are $\sim20\%$ more $B^0\to
K^+\pi^-$ than $\B0bar\to K^-\pi^+$ decays.

This measurement implies that after the "$K$-superweak" model~\cite{superweak},
now also "$B$-superweak" models are excluded.  I.e., models in which $CP$
violation only occurs in mixing are no longer viable.  This measurement also
establishes that there are sizable strong phases between the tree $(T)$ and
penguin $(P)$ amplitudes in charmless $B$ decays, since $|T/P|$ is estimated to
be not much larger than $|A_{K^-\pi^+}|$.  Such information on strong phases
will have broader implications for charmless nonleptonic decays and for
understanding the $B\to K\pi$ and $\pi\pi$ rates discussed in
Sec.~\ref{sec:pipi}.  

The bottom line is that, similar to $\epsilon'_K$, our theoretical understanding
at present is insufficient to either prove or rule out that the $CP$ asymmetry
in Eq.~(\ref{KpidirectCP}) is due to NP.

\subsection{CPV in mixing}

The two $B$ meson mass eigenstates are related to the flavor eigenstates via
\beq
|B_{L,H}\rangle = p |B^0\rangle \pm q |\B0bar\rangle\,.
\eeq
$CP$ is violated if the mass eigenstates are not equal to the $CP$ eigenstates.
This happens if $|q/p| \neq 1$, i.e., if the physical states are not
orthogonal, $\langle B_H | B_L\rangle \neq 0$, showing that this is an
intrinsically quantum mechanical phenomenon.

The simplest example of this type of $CP$ violation is the semileptonic decay
asymmetry to "wrong sign" leptons.  The measurements
give~\cite{Group(HFAG):2005rb}
\beq\label{ASL}
A_{\rm SL} = {\Gamma(\B0bar(t) \to \ell^+ X) - \Gamma(B^0(t) \to \ell^- X)
  \over \Gamma(\B0bar(t) \to \ell^+ X) + \Gamma(B^0(t) \to \ell^- X) }
= {1 - |q/p|^4 \over 1 + |q/p|^4} = -(3.0 \pm 7.8) \times 10^{-3} \,,
\eeq
implying $|q/p| = 1.0015 \pm 0.0039$, where the average is dominated by a recent
BELLE result~\cite{Nakano:2005jb}.  In semileptonic kaon decays the similar
asymmetry was measured~\cite{cplear}, in agreement with the expectation that it
is equal to $4\,{\rm Re}\, \epsilon$.

The calculation of $A_{\rm SL}$ is possible from first principles only in the
$m_b \gg \lqcd$ limit, using an operator product expansion to evaluate the
relevant nonleptonic rates.  Last year the NLO QCD calculation was
completed~\cite{Beneke:2003az,Ciuchini:2003ww}, predicting $A_{\rm SL} = -(5.5
\pm 1.3)\times 10^{-4}$, where I averaged the central values and quoted the
larger of the two theory error estimates.  (The similar asymmetry in the $B_s$
sector is expected to be $\lambda^2$ smaller.)  Although the experimental error
in Eq.~(\ref{ASL}) is an order of magnitude larger than the SM expectation, this
measurement already constraints new physics~\cite{Laplace:2002ik}, as the
$m_c^2/m_b^2$ suppression of $A_{\rm SL}$ in the SM can be avoided by NP.

\subsection{CPV in the interference between decay with and without mixing: $B\to
\psi K_{S,L}$}
\label{sec:sin2b}

It is possible to obtain theoretically clean information on weak phases in  $B$
decays to certain $CP$ eigenstate final states.  The  interference phenomena
between $B^0\to f_{CP}$ and $B^0 \to \B0bar \to f_{CP}$ is described by
\beq
\lambda_{f_{CP}} = \frac qp\, \frac{\ov A_{f_{CP}}}{A_{f_{CP}}} 
  = \eta_{f_{CP}}\, \frac qp\, \frac{\ov{A}_{\ov{f}_{CP}}}{A_{f_{CP}}} \,,
\eeq
where $\eta_{f_{CP}} = \pm 1$ is the $CP$ eigenvalue of $f_{CP}$. 
Experimentally one can study the time dependent $CP$ asymmetry,
\beq\label{SCdef}
a_{f_{CP}} = {\Gamma[\B0bar(t)\to f] - \Gamma[B^0(t)\to f]\over
  \Gamma[\B0bar(t)\to f] + \Gamma[B^0(t)\to f] }\qquad \\
= S_{f_{CP}} \sin(\Delta m\, t) - C_{f_{CP}} \cos(\Delta m\, t)\,,\nn
\eeq
where
\beq
S_f = {2\,{\rm Im}\,\lambda_f\over
  1+|\lambda_f|^2}\,, \qquad
C_f (= - A_f) = {1-|\lambda_f|^2 \over 1+|\lambda_f|^2}\,.
\eeq
If amplitudes with one weak phase dominate a decay then $a_{f_{CP}}$ measures a
phase in the Lagrangian theoretically cleanly.  In this case $C_f=0$, and 
$S_{f_{CP}} = {\rm Im}\,\lambda_{f_{CP}} = \sin(\arg\lambda_{f_{CP}})$, where
${\rm arg}\lambda_{f_{CP}}$ is the phase difference between the $\B0bar \to f$
and $\B0bar \to B^0\to f$ decay paths.

The theoretically cleanest example of this type of $CP$ violation is $B\to \psi
K^0$.  While there are tree and penguin contributions to the decay with
different weak phases, the dominant part of the penguin amplitudes have the same
weak phase as the tree amplitude.  Therefore, contributions with the tree
amplitude's weak phase dominate, to an accuracy better than $\sim$1\%.  In the
usual phase convention $S_{\psi K_{S,L}} = \mp \sin[ (B\mbox{-mixing}=-2\beta) +
(\mbox{decay}=0)  + (K\mbox{-mixing}=0)]$, so we expect $S_{\psi K_{S,L}} =
\pm\sin2\beta$ and $C_{\psi K_{S,L}} = 0$ to a similar accuracy.  The current
world average is
\beq\label{s2b}
\sin2\beta = 0.687 \pm 0.032\,,
\eeq
which is now a 5\% measurement.  In the last two years the $2\beta$ vs.\
$\pi-2\beta$ discrete ambiguity has also been resolved by ingenious studies of
the time dependent angular analysis of $B\to \psi K^{*0}$ and the time dependent
Dalitz plot analysis of $B^0 \to \D0bar h^0$ with $\D0bar \to K_S\pi^+\pi^-$,
pioneered by BABAR~\cite{Verderi:2004jp} and BELLE~\cite{Bondar:2005gk},
respectively.  As a result, the negative $\cos 2\beta$ solutions are excluded,
eliminating two of the four discrete ambiguities.

\begin{figure}[t]
\centerline{\includegraphics[width=0.45\textwidth]{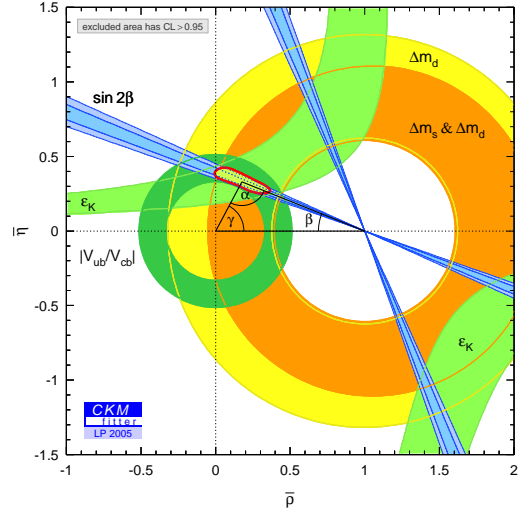}}
\caption{The present CKM fit using the measurements of $\epsilon_K$,
$|V_{ub}/V_{cb}|$, $\Delta m_{d,s}$, and $\sin2\beta$.}
\label{fig:smplot}
\end{figure}

To summarize, $S_{\psi K}$ was the first observation of $CP$ violation outside
the kaon sector, and the first observation of an ${\cal O}(1)$ $CP$ violating
effect.  It implies that models with approximate $CP$ symmetry (in the sense
that all CPV phases are small) are excluded.  The constraints on the CKM matrix
from the measurements of $S_{\psi K}$, $|V_{ub}/V_{cb}|$, $\epsilon_K$, $B$ and
$B_s$ mixing are shown in Fig.~\ref{fig:smplot} using the CKMfitter
package~\cite{ckmfitter,Charles:2004jd}.  The results throughout this paper are
based on the latest averages, except for $|V_{ub}|$, for which the
pre-Lepton-Photon 2005 value is used, $|V_{ub}| = (4.05 \pm 0.13 \pm 0.50)
\times 10^{-3}$, as explained in Sec.~\ref{sec:Vub}.  The overall consistency
between these measurements was the first precise test of the CKM picture.  It
also implies that it is unlikely that ${\cal O}(1)$ deviations from the SM can
occur, and one should look for corrections rather than alternatives of the CKM
picture.

\subsection{Other $CP$ asymmetries that are approximately $\sin2\beta$ in the
SM}
\label{sec:Bbs}

The $b\to s$ transitions, such as $\B0bar\to \phi K$, $\eta' K$, $K^+ K^- K_S$,
etc., are dominated by one-loop (penguin) diagrams in the SM, and therefore new
physics could compete with the SM contributions~\cite{Grossman:1996ke}.  Using
CKM unitarity we can write the contributions to such decays as a term
proportional to $V_{cb} V_{cs}^*$ and another proportional to $V_{ub}
V_{us}^*$.  Since their ratio is about 0.02, we expect amplitudes with the
$V_{cb} V_{cs}^*$ weak phase to dominate these decays as well.  Thus, in the SM,
the measurements of $-\eta_f S_f$ should agree with $S_{\psi K}$ (and $C_f$
should vanish) to an accuracy of order $\lambda^2 \sim 0.05$.  

If the SM and NP contributions are both significant, the $CP$ asymmetries depend
on their relative size and phase, which depend on hadronic matrix elements. 
Since these are mode-dependent, the asymmetries will, in general, be different
between the various modes, and different from $S_{\psi K}$.  One may also find
$C_f$ substantially different from 0.

\begin{table}[t]
\centerline{\begin{tabular}{|ccccc|}  
\hline
Dominant  &  \raisebox{-6pt}[0pt][-6pt]{$f_{CP}$}  
  &  SM allowed range of$^{\,*}$ \hspace*{-.3cm}
  &  \raisebox{-6pt}[0pt][-6pt]{$-\eta_f S_f$}
  &  \raisebox{-6pt}[0pt][-6pt]{$C_f$} 
\\[-2pt]
process  & &  $|\!-\eta_{f_{CP}} S_{f_{CP}} - \sin2\beta|$
& & \\ \hline\hline
$b\to c \bar c s$  &  $\psi K^0$  &  $<0.01$  
  &  $+0.687 \pm 0.032$  &  $+0.016\pm0.046$
\\ \hline
$b\to c \bar c d$  &  $\psi \pi^0$  &  $\sim 0.2$
  &  $+0.69 \pm 0.25$  &  $-0.11\pm0.20$
\\ 
  &  $D^{*+}D^{*-}$  &  $\sim 0.2$
  &  $+0.67 \pm 0.25$  &  $+0.09\pm0.12$
\\ 
  &  $D^+D^-$  &  $\sim 0.2$
  &  $+0.29 \pm 0.63$  &  $+0.11\pm0.36$
\\ \hline
$b\to s \bar q q$  &  $\phi K^0$  &  $< 0.05$
  &  $+0.47 \pm 0.19$  &  $-0.09\pm0.14$
\\ 
  &  $\eta' K^0$  &  $< 0.05$
  &  $+0.48 \pm 0.09$  &  $-0.08\pm0.07$
\\   
  &  $\pi^0 K_S$  &  $\sim 0.15$
  &  $+0.31 \pm 0.26$  &  $-0.02\pm0.13$
\\ 
  &  $K^+ K^- K_S$  &  $\sim 0.15$
  &  $+0.51 \pm 0.17$  &  $+0.15\pm0.09$
\\ 
  &  $K_S K_S K_S$  &  $\sim 0.15$
  &  $+0.61 \pm 0.23$  &  $-0.31\pm0.17$
\\
  &  $f^0 K_S$  &  $\sim 0.25$
  &  $+0.75 \pm 0.24$  &  $+0.06\pm0.21$ 
\\
  &  $\omega K_S$  &  $\sim 0.25$
  &  $+0.63 \pm 0.30$  &  $-0.44\pm0.23$ 
\\ \hline
\end{tabular}}
\caption{$CP$ asymmetries for which the SM predicts $-\eta_f S_f \approx
\sin2\beta$.  The 3rd column contains my estimates of limits on the deviations
from $\sin2\beta$ in the SM (strict bounds are worse), and the last two columns
show the world averages~\cite{Group(HFAG):2005rb}. (The $CP$-even fractions in
$K^+ K^- K_S$ and $D^{*+}D^{*-}$ are determined experimentally.)}
\label{tab:exp_fCP}
\end{table}

The averages of the latest BABAR and BELLE results are shown in
Table~\ref{tab:exp_fCP}.  The two data sets are fairly consistent by now.  The
largest hint of a deviation from the SM is now in the $\eta' K$ mode,
\beq\label{etapdiff}
S_{\psi K} - S_{\eta'K} =  0.21 \pm 0.10\,, \qquad
S_{\psi K} - S_{\phi K} =  0.22 \pm 0.19\,,
\eeq
which is $2\sigma$.  The average $CP$ asymmetry in all $b\to s$ modes, which
also equals $S_{\psi K}$ in the SM, has a bit more significant deviation,
$S_{\psi K} - \langle -\eta_f S_{f(b\to s)} \rangle = 0.18\pm0.07$. This is
currently a $2.6\sigma$ effect, however, this average is not too meaningful,
because some of the modes included may deviate significantly from $S_{\psi K}$
in the SM.  The third column in Table~\ref{tab:exp_fCP} shows my estimates of
limits on the deviations from $S_{\psi K}$ in the SM.  The hadronic matrix
elements multiplying the generic ${\cal O}(0.05)$ suppression of the "SM
pollution" are hard to bound model independently~\cite{Grossman:2003qp}, so
strict bounds are weaker, while model calculations obtain better limits.

To understand the significance of these measurements, note that a very
conservative bound using $SU(3)$ flavor symmetry using the current experimental
limits on related modes gives~\cite{Grossman:2003qp,Chiang:2003rb} $|S_{\psi K}
- S_{\eta'K}| < 0.2$ in the SM.  Estimates based on
factorization~\cite{Buchalla:2005us} obtain deviations at the 0.02 level.  Thus,
$|S_{\psi K} - S_{\eta'K}| \approx 0.2$ would be a sign of NP if established at
the $5\sigma$ level.  (The deviation of $S_{\phi K}$ from $S_{\psi K}$ is now
statistically insignificant, but the present central value with a smaller error
could still establish NP.)  Such a discovery would exclude in addition to the
SM, models with minimal flavor violation, and universal SUSY models, such as
gauge mediated SUSY breaking.

\section{Measurements of $\alpha$ and $\gamma$}
\label{sec:alphagamma}

To clarify terminology, I'll call a measurement of $\gamma$ the determination of
the phase difference between $b\to u$ and $b\to c$ transitions, while $\alpha
(\equiv \pi - \beta - \gamma)$ will refer to the measurements of $\gamma$ in the
presence of $B-\Bbar$ mixing.   The main difference between the measurements of
$\gamma$ and those of the other two angles is that $\gamma$ is measured in
entirely tree-level processes, so it is unlikely that new physics could modify
it.  It is therefore very important in searching for and constraining new
physics.  Interestingly, the best methods for measuring both $\alpha$ and
$\gamma$ are new since 2003.

\subsection{$\alpha$ from $B\to \pi\pi$, $\rho\rho$ and $\rho\pi$}
\label{sec:alpha}

\begin{figure}[t]
\centerline{\includegraphics[width=.45\textwidth]{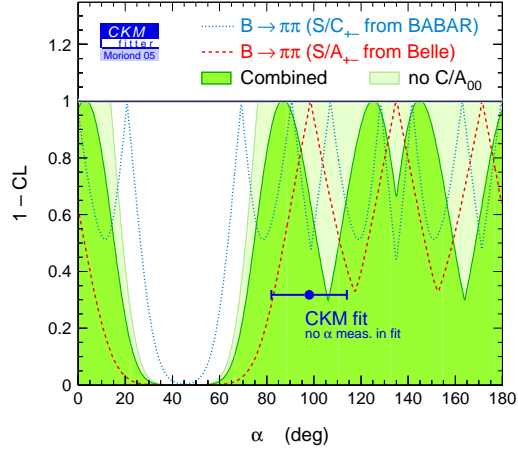}}
\caption{Constraints on $\alpha$ from $B\to\pi\pi$.  Some of the eight mirror
solutions may overlap.}
\label{fig:alphapipi}
\end{figure}

In contrast to $B\to \psi K$, which is dominated by amplitudes with one weak
phase, in $B\to \pi^+\pi^-$ there are two comparable contributions with
different weak phases.  Therefore, to determine $\alpha$ model independently, it
is necessary to carry out an isospin analysis~\cite{Gronau:1990ka} (for other
possibilities, see Sec.~\ref{sec:pipi}).  The hardest ingredients are the
measurement of the $\pi^0\pi^0$ rate,
\beq\label{pi0pi0}
{\cal B}(B\to \pi^0\pi^0) = (1.45 \pm 0.29) \times 10^{-6}\,,
\eeq
and the $CP$-asymmetry,
\beq\label{C00}
{\Gamma(\Bbar\to \pi^0\pi^0) - \Gamma(B\to \pi^0\pi^0)
  \over \Gamma(\Bbar\to \pi^0\pi^0) + \Gamma(B\to \pi^0\pi^0)}
  = 0.28 \pm 0.39.
\eeq
If these measurements were precise, one could pin down from the isospin analysis
the penguin pollution, $\Delta\alpha \equiv \alpha - \alpha_{\rm eff}$
($2\alpha_{\rm eff} = {\rm arg}\, \lambda_{\pi^+\pi^-} = \arcsin [S_{\pi^+\pi^-}
/(1-C_{\pi^+\pi^-}^2)^{1/2}]$).  In Fig.~\ref{fig:alphapipi}, the dark shaded
region shows the confidence level using Eq.~(\ref{C00}), while the light shaded
region is the constraint without it.  One finds $|\Delta\alpha| < 37^\circ$ at
90\% CL, a small improvement over the $39^\circ$ bound without Eq.~(\ref{C00}). 
This indicates that it will take a lot more data to determine $\alpha$
precisely.  In addition, the BABAR~\cite{Aubert:2004aq} and
BELLE~\cite{Abe:2004mp} results are still not quite consistent; see
Table~\ref{tab:pipi}.

\begin{table}[b]
\centerline{\begin{tabular}{|c|cc|}  
\hline
$B\to\pi^+\pi^-$  &  $S_{\pi^+\pi^-}$  &  $C_{\pi^+\pi^-}$ 
\\ \hline\hline
BABAR  &  $-0.30 \pm 0.17$  &  $-0.09 \pm 0.15$ \\
BELLE  &  $-0.67 \pm 0.17$  &  $-0.56 \pm 0.13$ \\
average  &  $-0.50 \pm 0.12\ [0.18]$  &  $-0.37 \pm 0.10\ [0.23]$ 
\\ \hline
\end{tabular}}
\caption{$CP$ asymmetries in $B\to \pi^+\pi^-$.  The brackets show the errors
inflated using the PDG perscription.}
\label{tab:pipi}
\end{table}

The $B\to \rho\rho$ mode is more complicated than $B\to \pi\pi$, because a
vector-vector ($VV$) final state is a mixture of $CP$-even ($L=0$ and 2) and
-odd ($L=1$) components.  The $B\to \pi\pi$ isospin analysis applies for each
$L$ in $B\to\rho\rho$ (or for each transversity, and, therefore, for the
longitudinal polarization component as well).  The situation is simplified
dramatically by the experimental observation that in the $\rho^+\rho^-$ and
$\rho^+\rho^0$ modes the longitudinal polarization fraction is near unity, so
the $CP$-even fraction dominates.  Thus, one can simply bound $\Delta\alpha$
from~\cite{Aubert:2004wb}
\beq
{\cal B}(B\to \rho^0\rho^0) < 1.1 \times 10^{-6}\ (90\%\ {\rm CL})\,.
\eeq
The smallness of this rate implies that $\Delta\alpha$ in $B\to \rho\rho$ is
much smaller than in $B\to \pi\pi$.  To appreciate the difference, note that
${{\cal B}(B\to \pi^0\pi^0) / {\cal B}(B\to \pi^+\pi^0)} = 0.26 \pm 0.06$, while
${\cal B}(B\to \rho^0\rho^0) / {\cal B}(B\to \rho^+\rho^0) < 0.04\ \ (90\%\ {\rm
CL})$.  From $S_{\rho^+\rho^-}$ and the isospin bound on $\Delta\alpha$ one
obtains
\beq
\alpha = 96 \pm 10 \pm 4 \pm 11^\circ(\alpha-\alpha_{\rm eff})\,.
\eeq
Ultimately the isospin analysis is more complicated in $B\to\rho\rho$ than in
$\pi\pi$, because the finite width of the $\rho$ allows for the final state
to be in an isospin-1 state~\cite{Falk:2003uq}.  This only affects the results
at the ${\cal O}(\Gamma_\rho^2/m_\rho^2)$ level, which is smaller than other
errors at present.  With higher statistics, it will be possible to constrain
this effect using the data~\cite{Falk:2003uq}.

Finally, in $B\to\rho\pi$ it is possible in principle to use a Dalitz plot
analysis~\cite{Snyder:1993mx} of the interference regions of the
$\pi^+\pi^-\pi^0$ final state to obtain a model independent determination of
$\alpha$, without discrete ambiguities.  The first such analysis
gives~\cite{Aubert:2004iu}
\beq
\alpha = (113 ^{+27}_{-17} \pm 6)^\circ .
\eeq
Viewing $B\to \rho\pi$ as two-body decays, isospin symmetry gives two pentagon
relations~\cite{Lipkin:1991st}.  Solving them would require measurements of the
rates and $CP$ asymmetries in all the $B\to \rho^+\pi^-$,  $\rho^-\pi^+$, and
$\rho^0\pi^0$ modes, which are not available.  BABAR and BELLE agree on the
direct $CP$ asymmetries, and their average
\beq
A_{\pi^-\rho^+} = -0.47 ^{+0.13}_{-0.14}\,, \qquad
A_{\pi^+\rho^-} = -0.15 \pm 0.09\,,
\eeq
is a $3.6\sigma$ signal of direct $CP$ violation, i.e., $(A_{\pi^-\rho^+}$,
$A_{\pi^+\rho^-}) \neq (0,0)$.  Translating the available measurements to a
value of  $\alpha$ involves assumptions about factorization and $SU(3)$ flavor
symmetry, and are theory error dominated.

\begin{figure}[t]
\centerline{\includegraphics[width=.45\textwidth]{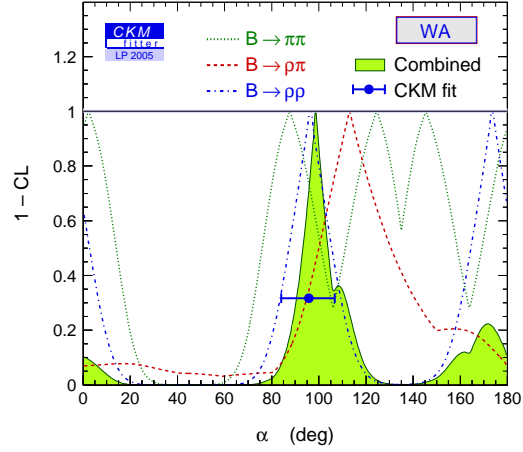}}
\caption{Confidence levels of the $\alpha$ measurements.}
\label{fig:alpha}
\end{figure}

Combining the $\rho\rho$ and $\pi\pi$ isospin analyses with the $\rho\pi$ Dalitz
plot analysis yields~\cite{Charles:2004jd}
\beq\label{alpha}
\alpha = (99^{+12}_{-9})^\circ ,
\eeq
which is shown in Fig.~\ref{fig:alpha}.  This direct determination of $\alpha$
is already more precise than it is from the CKM fit (without using $\alpha$ and
$\gamma$), which gives $\alpha = (98 \pm 16)^\circ$.

\subsection{$\gamma$ from $B^\pm\to DK^\pm$}

Here the idea is to measure the interference of $B^- \to D^0 K^-$ ($b\to c\bar u
s$) and $B^-\to \D0bar K^-$ ($b \to u \bar c s$) transitions, which can be
studied in final states accessible in both $D^0$ and $\D0bar$ decays. The key is
to extract the $B$ and $D$ decay amplitudes, the relative strong phases, and the
weak phase $\gamma$ from the data.  A practical complication is that the
precision depends sensitively on the ratio of the interfering amplitudes,
\beq
r_B \equiv {A(B^-\to \D0bar K^-) \over A(B^-\to D^0 K^-)}\,,
\eeq
which is around $0.1-0.2$.  In the original GLW method~\cite{GL,GW} one
considers $D$ decays to $CP$ eigenstate final states, such as $B^\pm\to
D^{(*)}_{CP} K^{\pm(*)}$.  To overcome the smallness of $r_B$ and make the
product of the $B$ and $D$ decay amplitudes comparable in magnitude, the ADS
method~\cite{ADS} considers final states where Cabibbo-allowed $\D0bar$ and
double Cabibbo-suppressed $D^0$ decays interfere.  So far the constraints on
$\gamma$ from these analyses are fairly weak.  There are other possibilities;
e.g., if $r_B$ is not much below $\sim0.2$ then studying single
Cabibbo-suppressed $D\to KK^*$ decays may be
advantageous~\cite{Grossman:2002aq}, or in three-body $B$ decays the color
suppression can be avoided~\cite{APS}.

It was recently realized~\cite{bondar,Giri:2003ty} that both $D^0$ and $\D0bar$
have Cabibbo-allowed decays to certain three-body final states, such as
$K_S\pi^+\pi^-$.  This analysis can be optimized by studying the Dalitz plot
dependence of the interference, and there is only a two-fold discrete
ambiguity.  The best present determination of $\gamma$ comes from this
analysis.  BELLE~\cite{Abe:2004gu} and BABAR~\cite{Aubert:2005yj} obtained
\beq
\gamma = 68^{+14}_{-15} \pm 13 \pm 11^\circ\,, \qquad
\gamma = 67 \pm 28 \pm 13 \pm 11^\circ\,,
\eeq
where the last uncertainty is due to the $D$ decay modelling.  The error is very
sensitive to the central value of the amplitude ratio $r_B$ (and $r_B^*$ for the
$D^*K$ channel), for which BELLE found somewhat larger central values than
BABAR.  The same values of $r_B^{(*)}$ also enter the ADS analyses, and the data
can be combined to fit for $\gamma$.  Combining the GLW, ADS, and Dalitz
analyses, one finds~\cite{Charles:2004jd}
\beq\label{gamma}
\gamma = \big( 63^{+15}_{-12} \big)^\circ . 
\eeq
More data will reduce the error of $\gamma$, allow for a significant measurement
of $r_B$, and reduce the dependence on the $D$ decay modelling.

\section{Implications of the $\alpha$ and $\gamma$ measurements}
\label{sec:NPmix}

\begin{figure}[t]
\centerline{\includegraphics[width=.49\textwidth]{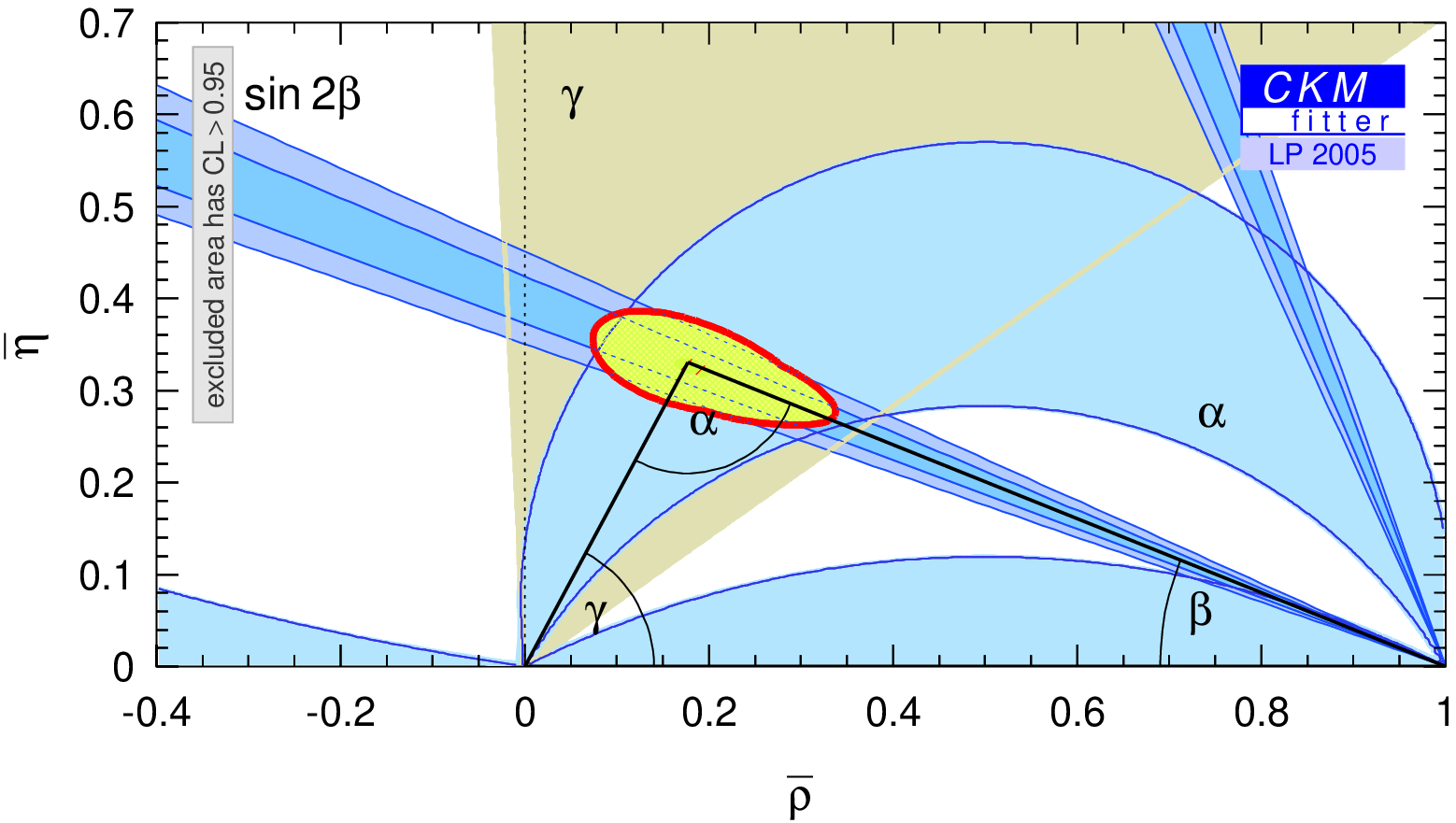}\hfill
\includegraphics[width=.49\textwidth]{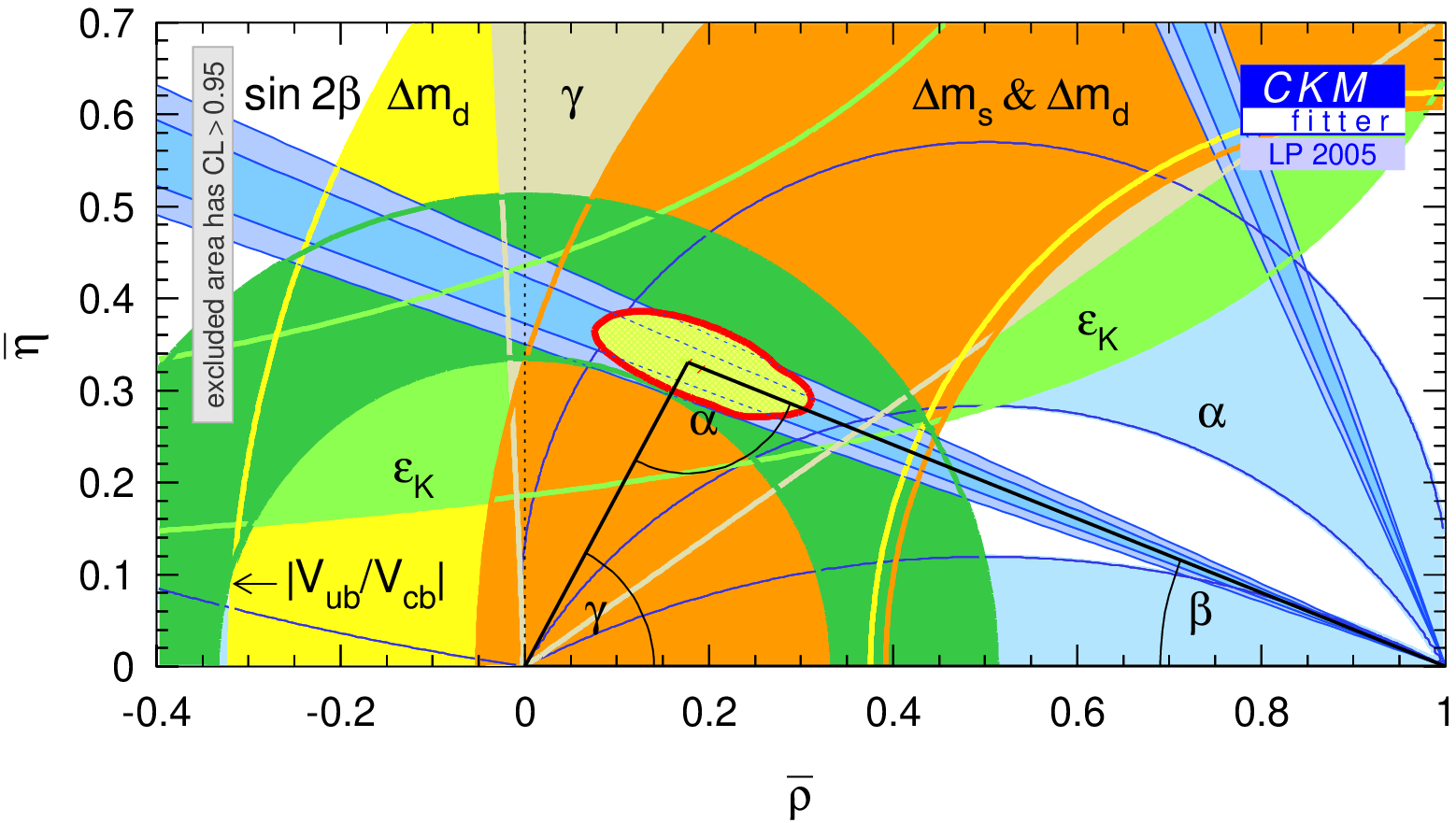}}
\caption{SM CKM fit including only the angle measurements (left), and all
available measurements not limited by theoretical uncertainties (right).}
\label{fig:smplotnew}
\end{figure}

Since the goal of the $B$ factories is to overconstrain the CKM matrix, one
should include in the CKM fit all measurements that are not limited by
theoretical uncertainties.  The result of such a fit is shown in the right plot
in Fig.~\ref{fig:smplotnew}, which includes in addition to the inputs in
Fig.~\ref{fig:smplot} the above measurements of $\alpha$ and $\gamma$.  The
left plot shows the fit using the angle measurements only, and indicates that
determination of $\rhobar,\etabar$ from the angles alone is almost as precise as
from all inputs combined.  The allowed region of $\rhobar,\etabar$ shrinks only
slightly compared to Fig.~\ref{fig:smplot}, and the most interesting implication
of the $\alpha$ and $\gamma$ measurements is the reduction in the allowed range
of the $B_s-\Bbar_s$ mixing frequency.  While the fit in Fig.~\ref{fig:smplot}
gives $\Delta m_s = \big(17.9 ^{+10.5} _{-1.7}\, {}^{[+20.0]}_{[-2.8]}\big)\,
{\rm ps}^{-1}$ at $1\sigma\ [2\sigma]$, the full fit gives $\Delta m_s =
\big(17.9 ^{+3.6} _{-1.4}\, {}^{[+8.6]}_{[-2.4]}\big)\, {\rm ps}^{-1}$.

\subsection{New physics in $B^0-\B0bar$ mixing}

\begin{figure}[tp]
\centerline{\hspace{8pt}
\includegraphics[width=.47\textwidth,height=.44\textwidth]{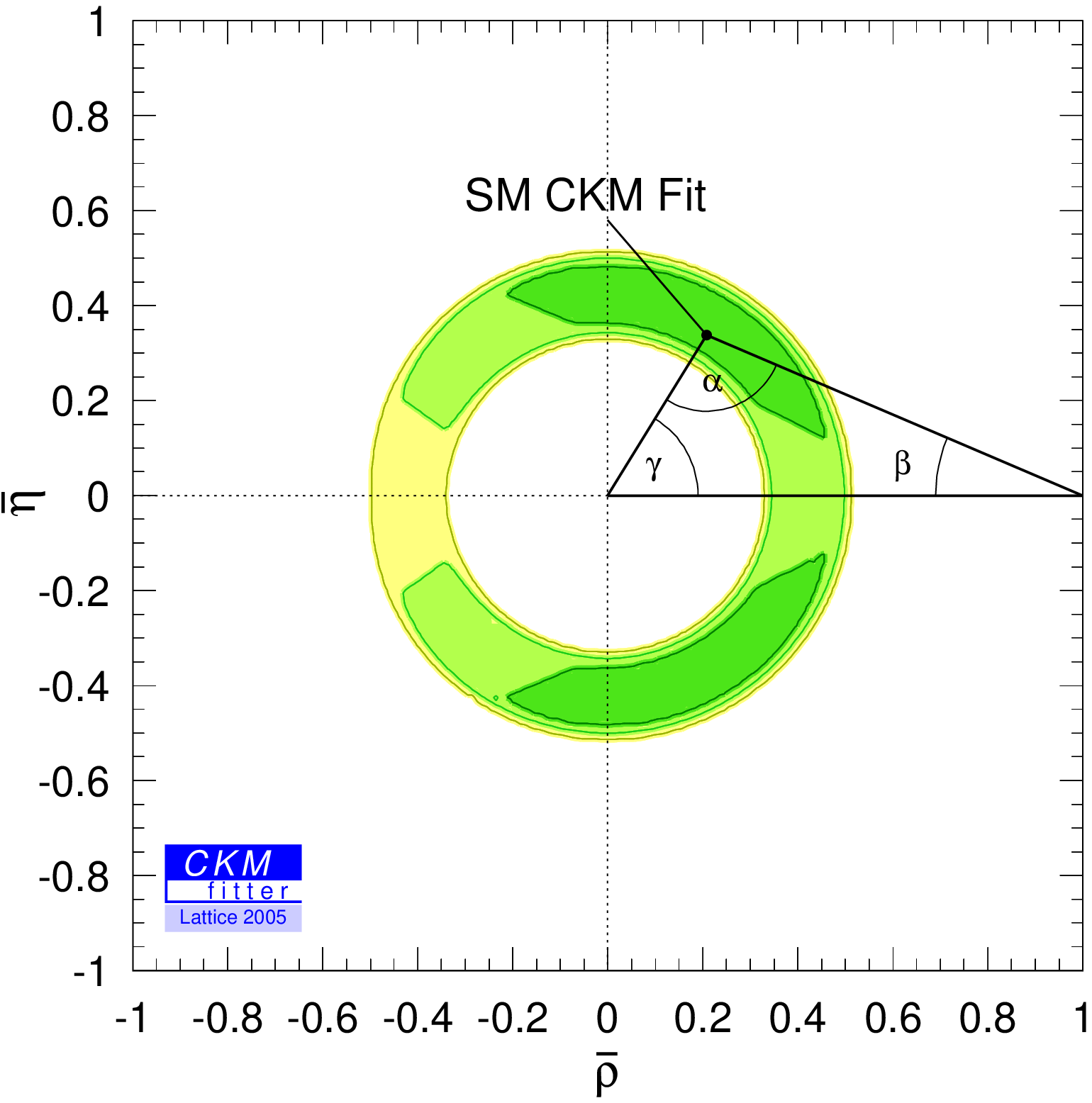}
\hspace{6pt}
\includegraphics[width=.47\textwidth,height=.44\textwidth]{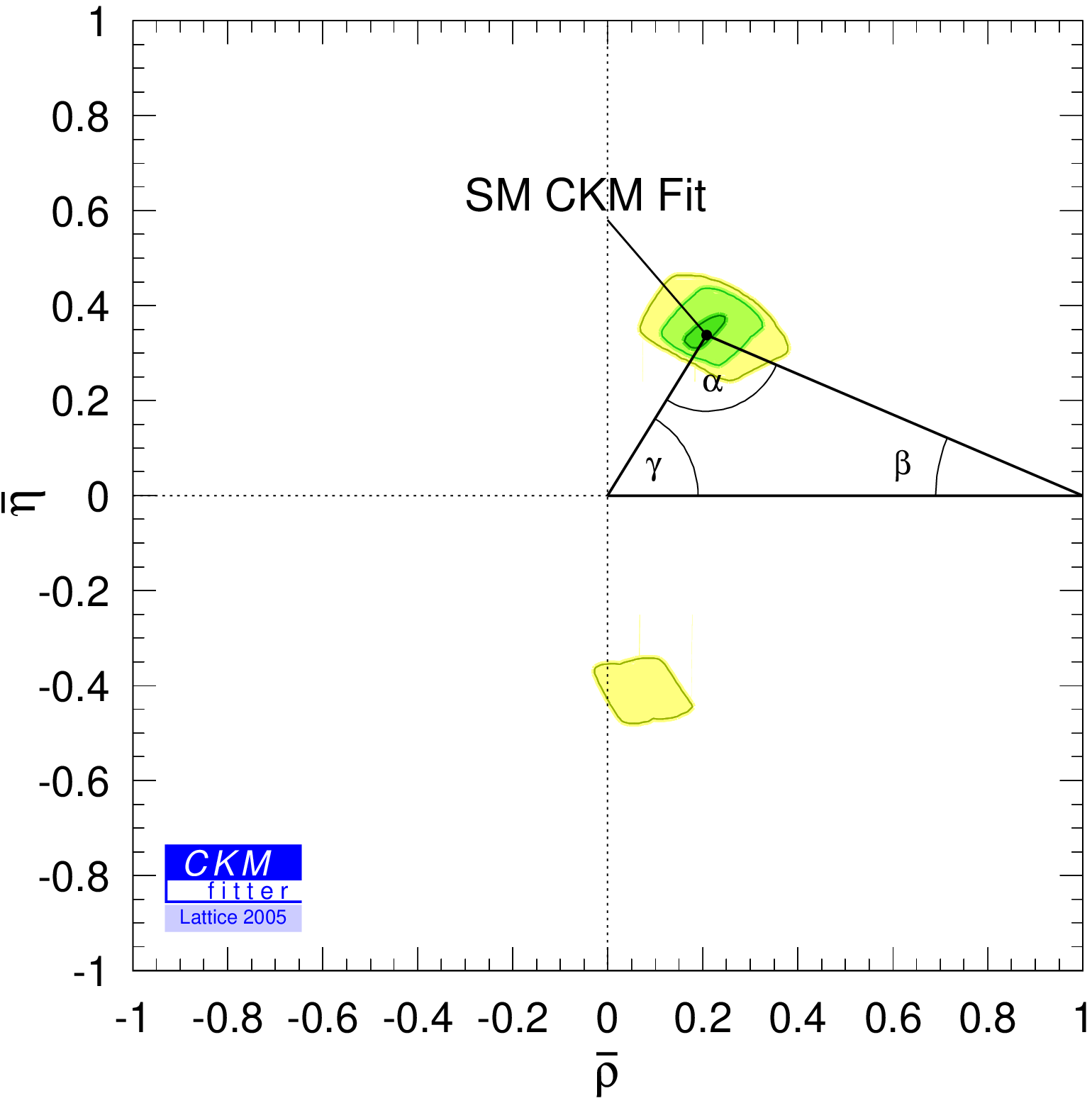}}
\vspace*{4pt}
\centerline{\includegraphics[width=.49\textwidth]{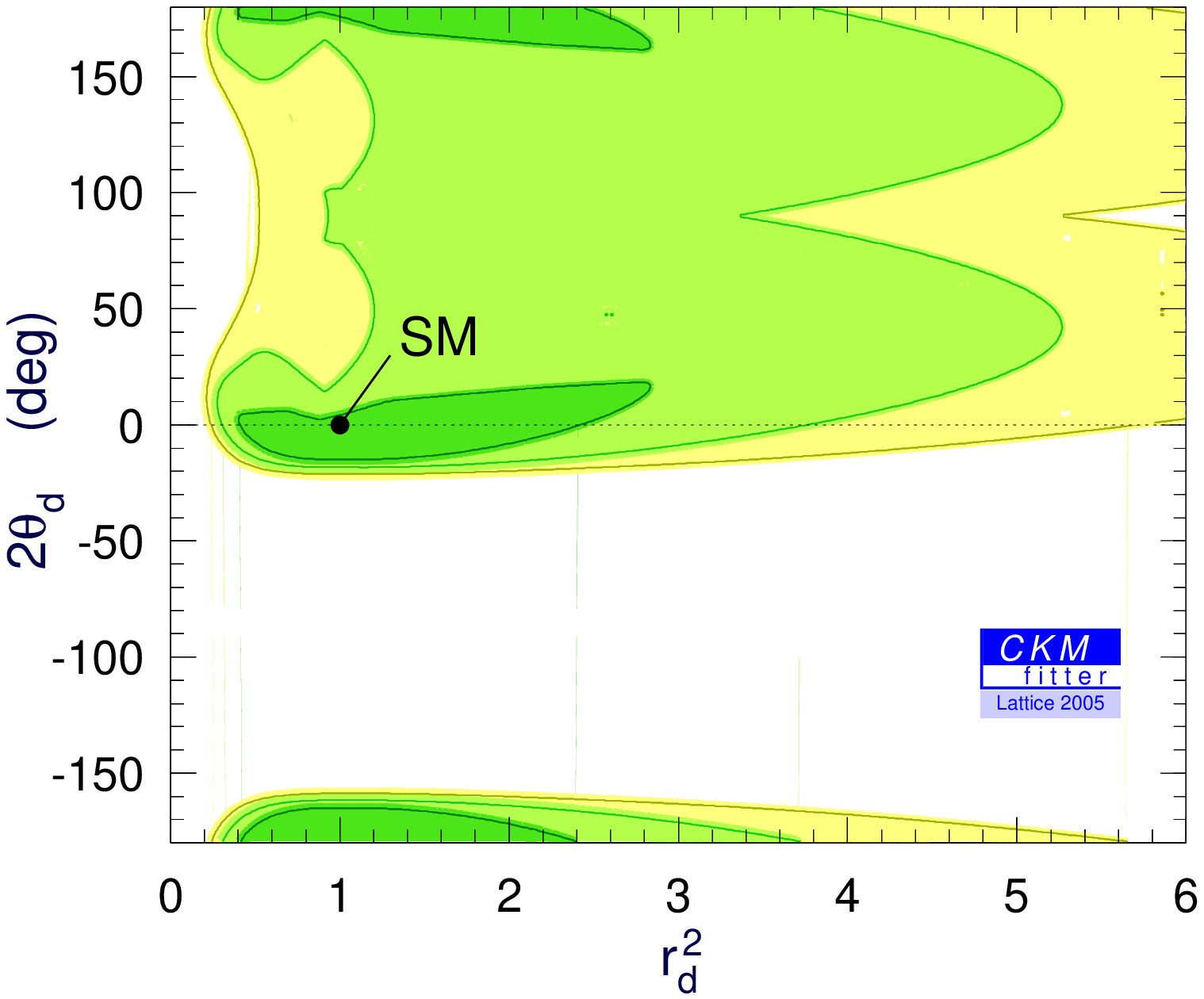}
\includegraphics[width=.49\textwidth]{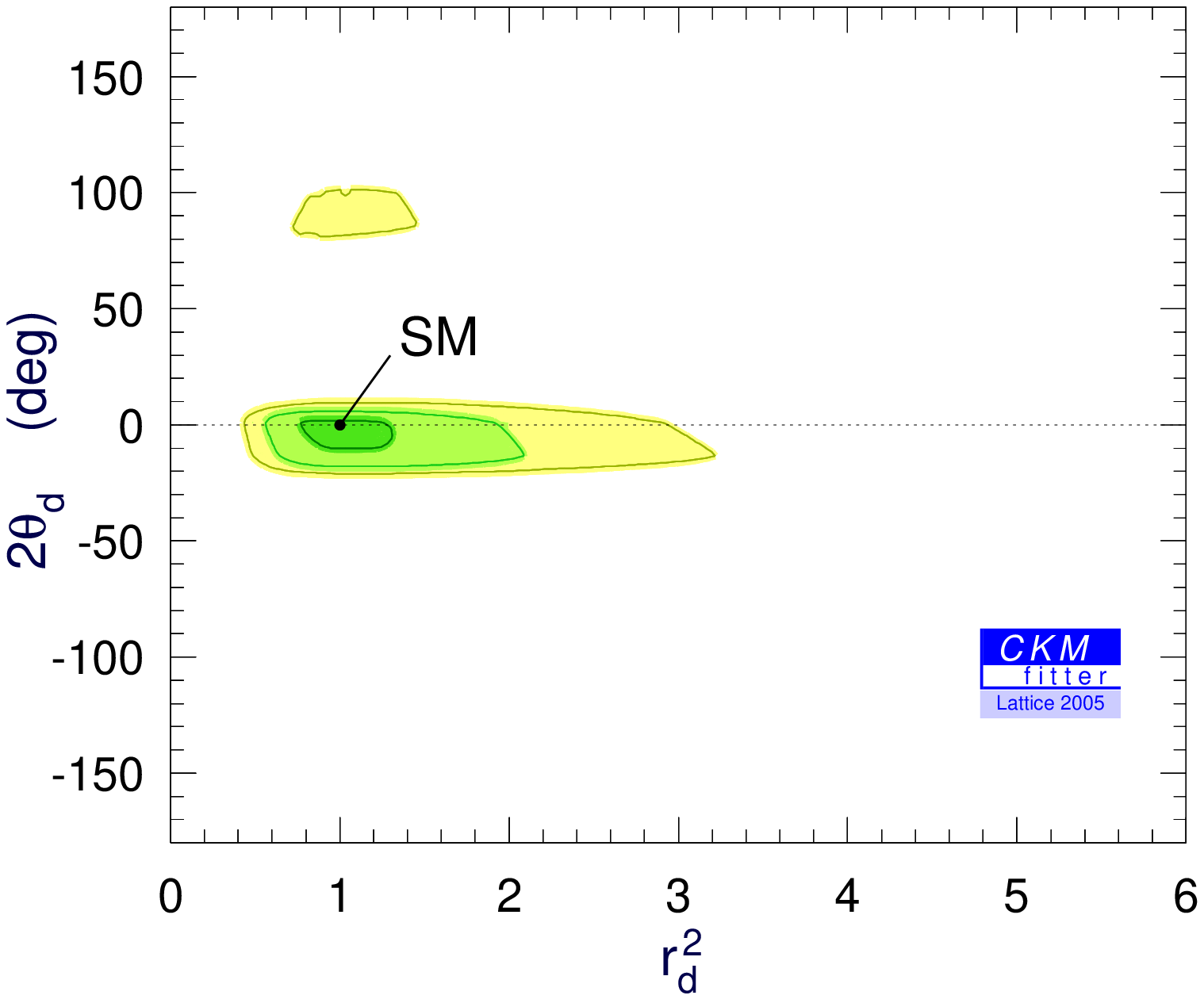}}
\caption{Allowed regions in the $\rhobar-\etabar$ plane (top) and the
$r_d^2-2\theta_d$ plane (bottom) in the presence of new physics in $B-\Bbar$
mixing.  The left [right] plots are the allowed regions without [with] the
constraints on $\alpha$, $\gamma$, and $\cos2\beta$.  The dark, medium, and
light shaded areas have CL $>$ 0.90, 0.32, and 0.05, respectively.}
\label{fig:npplot}

\vspace{.45cm}
\centerline{\includegraphics[width=.5\textwidth]{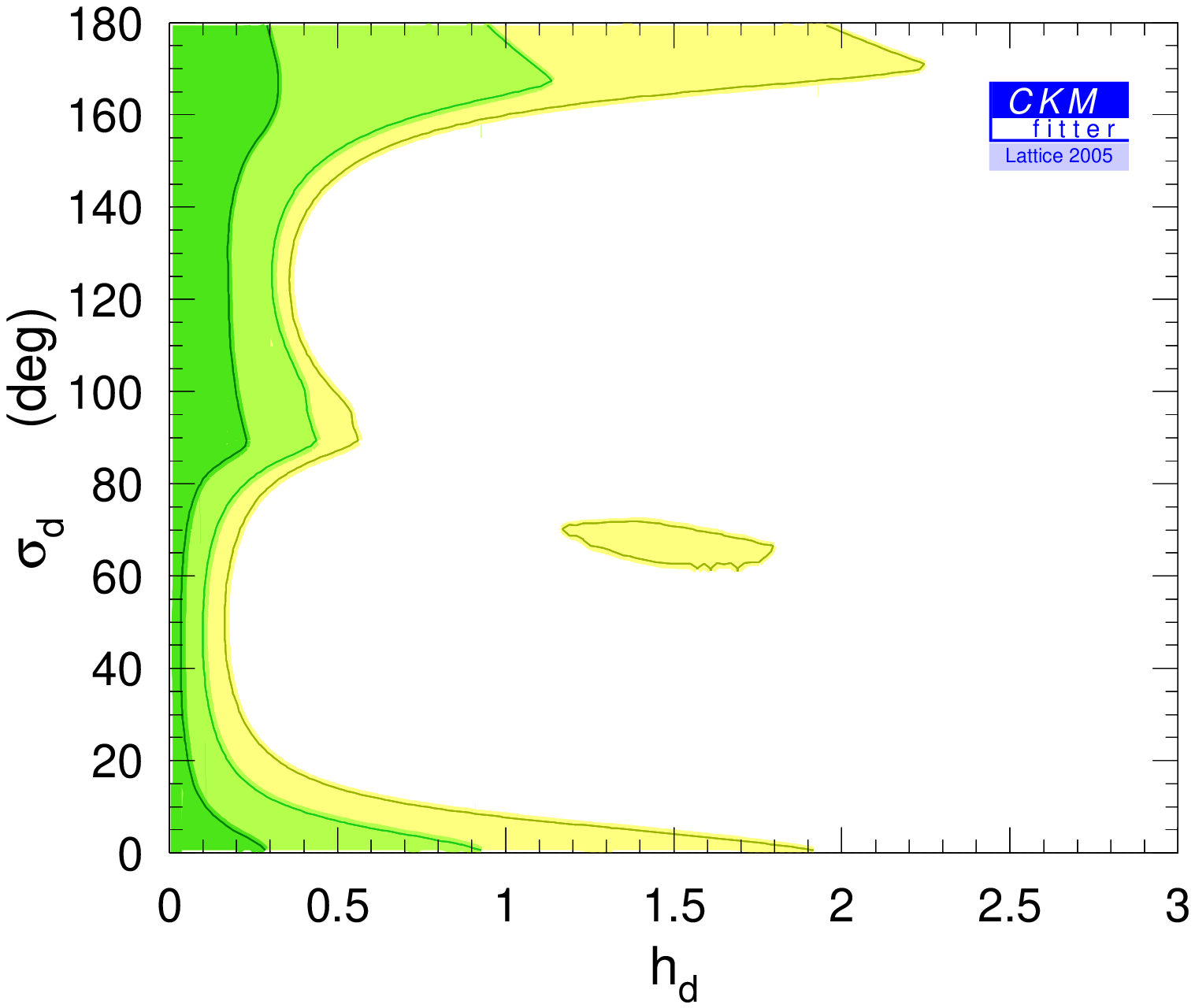}}
\caption{Allowed regions in the $h_d-\sigma_d$ plane.}
\label{fig:hdsd}
\end{figure}

In a large class of models the dominant NP effect in $B$ physics is to modify
the $B^0-\B0bar$ mixing amplitude~\cite{Soares:1992xi}, which can be
parameterized as 
\beq
M_{12} = M_{12}^{\rm (SM)} r_d^2\, e^{2i\theta_d} 
  = M_{12}^{\rm (SM)} (1 + h_d\, e^{2i\sigma_d}).
\eeq
Then $\Delta m_B = r_d^2\, \Delta m_B^{\rm (SM)}$, $S_{\psi K} = \sin(2\beta +
2\theta_d)$,  $S_{\rho^+\rho^-} = \sin(2\alpha - 2\theta_d)$, while the
tree-level measurements $|V_{ub}/V_{cb}|$ and $\gamma$ extracted from $B\to DK$
are unaffected.  Since $\theta_d$ drops out from $\alpha+\beta$, the
measurements of $\alpha$, together with $\beta$, are effectively equivalent in
these models to NP-independent measurements of $\gamma$ (up to discrete
ambiguities).  Measurements irrelevant for the SM CKM fit, such as the $CP$
asymmetry in semileptonic decays, $A_{\rm SL}$, become important for these
constraints~\cite{Laplace:2002ik}.

Figure~\ref{fig:npplot} shows the fit results using only $|V_{ub}/V_{cb}|$,
$\Delta m_B$, $S_{\psi K}$, and $A_{\rm SL}$ as inputs (left) and also including
the measurements of $\alpha$, $\gamma$, and $\cos2\beta$ (right) in the
$\rhobar-\etabar$ plane (top) and the $r_d^2-2\theta_d$ plane (bottom).  The
recent $\gamma$ and $\alpha$ measurements determine $\rhobar,\etabar$ from
(effectively) tree-level $B$ decays for the first time, independent of mixing,
and agree with the other SM
constraints~\cite{Ligeti:2004ak,Charles:2004jd,Bona:2004sj}.  The disfavored
"non-SM" region around $2\theta_d \sim 90^\circ$ is due to the $\eta<0$ region
in the top right plot and discrete ambiguities.  Thus, NP in $B^0-\B0bar$ mixing
is severely constrained for the first time.

The $h_d,\sigma_d$ parameterization is more convenient to study specific NP
scenarios, since in a given model it is an additive contribution to $M_{12}^{\rm
(SM)}$ that is directly calculable. The allowed range of $h_d,\sigma_d$ is shown
in Fig.~\ref{fig:hdsd}.  While the constraints are significant, new physics with
arbitrary weak phase may still contribute to $M_{12}$ at the level of $20\%$ of
the SM~\cite{Agashe:2005hk}.  Similar results for the constraints on NP in $K$
and $B_s$ mixing can be found in Refs.~\cite{Agashe:2005hk,Bona:2005eu}.  These
constraints would not follow from just measuring each CKM element one way, and
could be derived only due to overconstraining measurements.

\section*{Theoretical developments}

Studying $B$ decays is not only a window to new physics, it also allows us to
investigate the interplay of weak and strong interactions at a level of
unprecedented detail.  Many observables beyond those discussed so far are
sensitive to NP, and the question is in which cases we can disentangle signals
of NP from hadronic physics.

Most of the recent theoretical progress in understanding $B$ decays (using
continuum methods) utilize that $m_b$ is much larger than $\lqcd$.  In
particular, in the last few years there were significant developments toward a
model independent theory of certain exclusive semileptonic and nonleptonic
decays in the $m_B \gg \lqcd$ limit.  However, depending on the process under
consideration, the relevant hadronic scale may or may not be much smaller than
$m_b$ (and especially $m_c$).  For example, $f_\pi$, $m_\rho$, and $m_K^2/m_s$
are all of order $\lqcd$, but their numerical values span an order of
magnitude.  In most cases experimental guidance is needed to determine how well
the theory works.

\section{Inclusive semileptonic decays}

\subsection{$|V_{cb}|$ and $m_b$ from $B\to X_c\ell\bar\nu$}

I would like to use the determination of $|V_{cb}|$ from inclusive semileptonic
$B\to X_c\ell\bar\nu$ decay as an example to illustrate what we have learned
without lattice QCD (LQCD).  The state of the art is that using an operator
product expansion (OPE)~\cite{OPE} the semileptonic rate, as well as moments of
the lepton energy and the hadronic invariant mass spectra have been computed to
order $\lqcd^3/m_b^3$ and $\alpha_s^2\beta_0$.  The expressions are of the form
\beqa
&& \hspace*{-.75cm} 
\Gamma(B\to X_c\ell\bar\nu) = {G_F^2 |V_{cb}|^2\over 192\pi^3}\,
  \bigg({m_\Upsilon\over 2}\bigg)^{\!5}\, (0.534) \nn\\
\times\!\!\! &\!\!\bigg[&\!\!\! 1 - 0.22 \Big({\Lambda_{1S}\over 500\,\MeV}\Big)
  - 0.011 \Big({\Lambda_{1S}\over 500\,\MeV}\Big)^2
  - 0.052 \Big({\lambda_1\over (500\,\MeV)^2}\Big)
  - 0.071 \Big({\lambda_2\over (500\,\MeV)^2}\Big) \nn\\
&-& 0.006 \Big({\lambda_1\Lambda_{1S}\over (500\,\MeV)^3}\Big)
  + 0.011 \Big({\lambda_2\Lambda_{1S}\over (500\,\MeV)^3}\Big)
  - 0.006 \Big({\rho_1\over (500\,\MeV)^3}\Big)
  + 0.008 \Big({\rho_2\over (500\,\MeV)^3}\Big) \nn\\
&+& 0.011 \Big({T_1\over (500\,\MeV)^3}\Big)
  + 0.002 \Big({T_2\over (500\,\MeV)^3}\Big)
  - 0.017 \Big({T_3\over (500\,\MeV)^3}\Big)
  - 0.008 \Big({T_4\over (500\,\MeV)^3}\Big) \nn\\
&+& 0.096\epsilon - 0.030 \epsilon^2_{\rm BLM} 
  + 0.015 \epsilon \Big({\Lambda_{1S}\over 500\,\MeV}\Big) + \ldots \bigg] \,,
\eeqa
where $\Lambda_{1S} \equiv m_\Upsilon/2 - m_b^{1S}$ is related to a short
distance $b$ quark mass~\cite{Hoang:1998ng,Hoang:1999zc}, and the ${\cal
O}(\lqcd^2/m_b^2)$ corrections are parameterized by $\lambda_{1,2}$.  The other
terms are $\lqcd^3/m_b^3$ and perturbative corrections, where
$\varepsilon\equiv1$ counts the order and the BLM subscript denotes terms with
the highest power of $\beta_0$.

Such formulae are fitted to about $90$ observables.  The fits determine
$|V_{cb}|$ and the hadronic parameters, and their consistency provides a
powerful test of the theory.  The fits have been performed in several schemes
and give~\cite{Bauer:2004ve,Buchmuller:2005zv}, 
\beq
|V_{cb}| = (41.7 \pm 0.7)\times 10^{-3}\,,
\eeq
where I averaged the central values and kept the error quoted in each paper. 
For the quark masses one gets~\cite{Bauer:2004ve,Hoang:2005zw}
\beq
m_b^{1S} = (4.68\pm 0.03)\, \GeV\,, \qquad 
\ov m_c(\ov m_c) = (1.22\pm0.06)\, \GeV\,,
\eeq
which correspond to $\ov m_b(\ov m_b) = (4.18 \pm 0.04)\, \GeV$ and 
$m_c^{1S} = (1.41\pm 0.05)\, \GeV$.

\subsection{$B\to X_u\ell\bar\nu$, $X_s\gamma$ and $X_s\ell^+\ell^-$}
\label{sec:Vub}

One could easily spend a whole talk on inclusive heavy to light decays, but
since it has little relevance for LQCD, I will be brief.  The determination of
$|V_{ub}|$ is more complicated than that of $|V_{cb}|$, because of the large $B
\to X_c \ell \bar\nu$ background.  The total $B \to X_u \ell \bar\nu$ rate is
known at the 5\% level~\cite{Hoang:1998ng}, but the cuts used in most
experimental analyses to remove the $B\to X_c\ell\bar\nu$ background introduce
${\cal O}(1)$ dependence on a nonperturbative $b$ quark light-cone distribution
function (sometimes called the shape function).  At leading order, one universal
function occurs~\cite{Neubert:1993ch}, which can be extracted from $B\to
X_s\gamma$ and applied to the analyses of the measured $E_\ell$, $m_X$ or
$P_X^+(=E_X-|\vec p_X|)$ spectra in $B\to X_u \ell \bar\nu$.  At order
$\lqcd/m_b$ several new functions occur~\cite{Bauer:2001mh}, and it is not known
how to extract these from data.  The hadronic physics being parameterized by
functions is a significant complication compared to the determination of
$|V_{cb}|$, where it is encoded by numbers.

A different approach is to eliminate the $B \to X_c \ell \bar\nu$ background
using $q^2$ and $m_X$ cuts, in which case the local OPE remains
valid~\cite{Bauer:2000xf}.  The dependence on the shape function can also be
reduced by extending the measurements into the $B \to X_c \ell \bar\nu$ region.
Recent analyses could measure the $B\to X_u e \bar\nu$ rates for $p_e \ge
1.9$\,GeV, which is well below the charm threshold~\cite{Bornheim:2002du}.

\begin{figure}[t]
\centerline{\includegraphics[width=.5\textwidth]{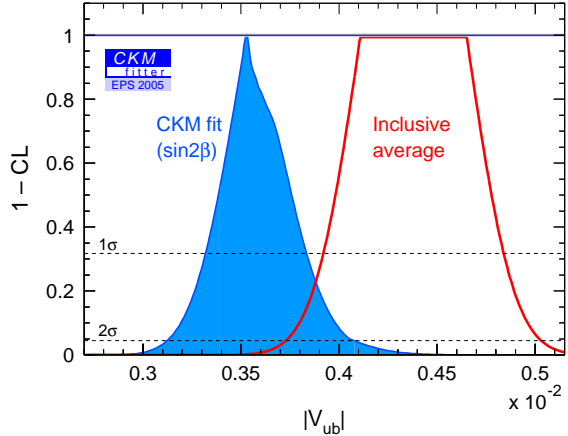}}
\caption{HFAG's inclusive average for $|V_{ub}|$ vs.\ the prediction from the
CKM fit.}
\label{fig:vubincl}
\end{figure}

Averaging the inclusive measurements, HFAG quotes $|V_{ub}| = (4.38 \pm 0.19 \pm
0.27) \times 10^{-3}$~\cite{Group(HFAG):2005rb}, which is in slight tension with
the CKM fit for $|V_{ub}|$ dominated by the $\sin2\beta$ measurement; see
Fig.~\ref{fig:vubincl}.  HFAG uses the prescription~\cite{Bosch:2004th}, and due
to concerns about how the shape function model dependence and error is
estimated, I use an older value of $|V_{ub}|$ (see end of Sec.~\ref{sec:sin2b}).

The loop-dominated $B\to X_s\gamma$ and $X_s\ell^+\ell^-$ decays received a lot
of attention, because of their sensitivity to new physics.  We recently learned
that both ${\cal B}(B\to X_s\gamma) =(3.39 ^{+0.30}_{-0.27}) \times 10^{-4}$
and  ${\cal B}(B\to X_s \ell^+\ell^-) = (4.5 \pm 1.0) \times
10^{-6}$~\cite{Group(HFAG):2005rb} agree with the SM at the 10\% and 20\% level,
respectively, which is a great triumph for the SM.  There is a  major ongoing
effort toward the NNLO calculation of ${\cal B}(B\to
X_s\gamma)$~\cite{Blokland:2005uk}, which may reduce the perturbation theory
error to $\lsim 5\%$.

As mentioned above, the $B\to X_s\gamma$ photon spectrum is also important for
the determination of the shape function that enters many $|V_{ub}|$
measurements.  It was realized recently that the same nonperturbative shape
function is also relevant for $B\to X_s\ell^+\ell^-$~\cite{Lee:2005pw}, where
the measured rate for $q^2 < m_\psi^2$ depends on it, because experimantally an
additional cut on $m_{X_s}$ has to be used.

These rare decay measurements may actually make model building more
interesting.  The present central values of the $CP$ asymmetries, $S_{\eta' K}$
and $S_{\phi K}$, can be reasonably accommodated by NP, such as SUSY (unlike
${\cal O}(1)$ deviations from $S_{\psi K}$ shown by the central values before
2004).  While $B\to X_s\gamma$ mainly constrains left-right ($LR$) mass
insertions in SUSY, $B\to X_s\ell^+\ell^-$ also constrains $RR$ and $LL$ mass
insertions contributing in penguin diagrams.  Thus, NP models have to satisfy a
growing number of interrelated constraints.

\section{Exclusive semileptonic heavy to light decays}

\subsection{What we knew a few years ago}

I will not talk about exclusive $B\to D^{(*)}\ell\bar\nu$ decays.  Its status
using LQCD was reviewed in the next talk~\cite{Okamoto:2005zg}.  In $B$ decays
to light mesons LQCD is also indispensible, as there is much more limited use of
heavy quark symmetry (HQS) than in $B\to D^{(*)}$.  Heavy quark spin symmetry
implies relations between the $B\to \rho\ell\bar\nu$, $K^*\ell^+\ell^-$, and
$K^*\gamma$ form factors in the large $q^2$ region~\cite{Isgur:1990kf}, but it
does not determine their normalization.  We shall return to the small $q^2$
region below.

To determine $|V_{ub}|$ with sub-10\% error from an exclusive decay, unquenched
LQCD calculations are required, which have started to become available, so far
limited to large $q^2$.  Without using LQCD, one can combine heavy quark and
chiral symmetries to form "Grinstein-type double
ratios"~\cite{Grinstein:1993ys}, whose deviation from unity is suppressed in
both symmetry limits.  For example,
\beq
{f_B\over f_{B_s}} \times {f_{D_s}\over f_D}
= 1 + {\cal O}\bigg( {m_s\over m_c}-{m_s\over m_b}\,,\;
  {m_s\over 1\,{\rm GeV}}\, {\alpha_s(m_c)-\alpha_s(m_b)\over\pi} \bigg) \,,
\eeq
and lattice calculations indicate that the deviation from unity is indeed at the
few  percent level.  Similar double ratios can be constructed for the
semileptonic form factors~\cite{Ligeti:1995yz,Ligeti:1997aq},
\beq
{f^{(B\to\rho\ell\bar\nu)} \over f^{(B\to K^*\ell^+\ell^-)}} \times 
{f^{(D\to K^*\ell\bar\nu)} \over f^{(D\to\rho\ell\bar\nu)}}\,,
\eeq
or for appropriately weighted $q^2$-spectra in these decays, and may be
experimentally  accessible soon.  Recently, the leading power corrections to the
HQS relations between the $B$ and $D$ decay form factors were
analyzed~\cite{BGDP}.  With data from LHCB and a super-$B$-factory the double
ratio~\cite{Ligeti:2003hp}
\beq
{{\cal B}(B\to\ell\bar\nu) \over {\cal B}(B_s\to \ell^+\ell^-)} \times 
{{\cal B}(D_s\to \ell\bar\nu) \over {\cal B}(D\to\ell\bar\nu)}
\eeq
could give a determination of $|V_{ub}|$ with theoretical errors at the few
percent level.

Interestingly, the most recent CLEO-c~\cite{Huang:2005iv} and
FOCUS~\cite{Link:2005xe} data are still consistent with no $SU(3)$ breaking in
the $D\to \rho\ell\bar\nu$ vs.\ $D\to K^*\ell\bar\nu$ form factors.  Averaging
these measurements gives ${\cal B}(D^+\to \rho^0\ell^+\nu) / {\cal B}(D^+\to
\K0bar{}^*\ell^+\nu) = 0.041 \pm 0.005$, while the theoretical prediction using
the measured $D\to K^*$ form factors and neglecting $SU(3)$ breaking in the
matrix elements is 0.044~\cite{Ligeti:1997aq}.

\subsection{A "one-page" introduction to SCET}
\label{sec:scet}

To discuss recent developments in understanding the semileptonic form factors at
small $q^2$ and some nonleptonic decays, we need to sketch some features of the
soft collinear effective theory (SCET)~\cite{sceta,scetb}.  It is a theory
designed to describe the interactions of energetic and low invariant mass
partons in the $Q \gg \Lambda$ limit.  SCET is constructed by introducing
distinct fields for the relevant degrees of freedom, and a power counting
parameter $\lambda$.  It is convenient to distinguish two theories, SCET$_{\rm
I}$ in which $\lambda=\sqrt{\lqcd/Q}$ and SCET$_{\rm II}$ in which
$\lambda=\lqcd/Q$~\cite{Bauer:2002aj}.  They are appropriate for final states
with invariant mass $Q\lambda$; i.e., SCET$_{\rm I}$ for jets and inclusive
$B\to X_s\gamma$, $X_u\ell\bar\nu$, $X_s\ell^+\ell^-$ decays in the shape
function regions ($m^2 \sim \lqcd Q$), and SCET$_{\rm II}$ for exclusive
hadronic final states ($m^2 \sim \lqcd^2$).  The fields in SCET are shown in
Table~\ref{tab:scet}.  It is convenient to use light-cone coordinates,
decomposing momenta as $p^\mu = \frac12 (\bar n\cdot p) n^\mu + p_\perp^\mu +
\frac12 (n\cdot p) \bar n^\mu$, where $n^2 = \bar n^2 = 0$ and $n \cdot \bar n =
2$.  For a light quark moving in the $n$ direction, $\psi(x) = \sum_p e^{-i
p\cdot x} \big[\frac14 \nslash\nbarslash \xi_{n,p}(x) + \frac14
\nbarslash\nslash \widetilde\xi_{n,p}(x)\big] $ separates the large
($\xi_{n,p}$) and small ($\widetilde\xi_{n,p}$) components of the spinor,
similar to HQET~\cite{Georgi:1990um} where $b(x) = \sum_v e^{-i m_b v\cdot x}
\big[\frac12 (1+\vslash) h_v^{(b)}\!(x) + \frac12 (1-\vslash) \widetilde
h_v^{(b)}\!(x)\big]$ separates the large ($h_v^{(b)}$) and small ($\widetilde
h_v^{(b)}$) components of a $b$ quark field.  Contrary to HQET, there is no
superselection rule in SCET, because collinear gluons can change $p$ without any
suppression.

\begin{table}
\centerline{\begin{tabular}{|cccc|} \hline
modes  &  fields  &  $p = (+, -, \perp)$  &  $p^2$ \\ \hline\hline
collinear  &  $\xi_{n,p},\ A_{n,q}^\mu$  &  $Q(\lambda^2,1,\lambda)$  
  &  $Q^2\lambda^2$\\[2pt]
soft  &  $q_q,\ A_s^\mu$  &  $Q(\lambda,\lambda,\lambda)$  &  $Q^2\lambda^2$
\\[2pt]
usoft  &  $q_{us},\ A_{us}^\mu$  &  $Q(\lambda^2,\lambda^2,\lambda^2)$  
  &  $Q^2\lambda^4$ \\ \hline
\end{tabular}}
\caption{Fields and their scalings in SCET.  There are no soft [ultrasoft]
fields in SCET$_{\rm I}$ [SCET$_{\rm II}$].}
\label{tab:scet}
\end{table}

In matching QCD on SCET$_{\rm I}$ (and when appropriate on SCET$_{\rm II}$) and
expanding the weak currents and the Lagrangian in $\lambda$, a complication is
that integrating out the off-shell degrees of freedom builds up Wilson lines. 
For example, the heavy to light current, $\bar q\, \Gamma\, b$, matches onto the
SCET current $\bar\xi_{n,p} W\, \Gamma\, h_v^{(b)}$, where $W$ is a Wilson line
built out of collinear gluons.  The theory also requires soft and ultrasoft
Wilson lines, usually denoted by $S$ and $Y$.  Powerful constraints on the
structure of SCET operators arise from the requirement of separate collinear,
soft, and usoft gauge invariances.  All fields transform under ultrasoft gauge
transformation, but, for example, heavy quark fields do not transform under
collinear ones.  Some of the simplifications in dealing with nonperturbative
phenomena in SCET$_{\rm I}$ arise from the observation that by the field
redefinitions~\cite{scetb}
\beq
\xi_{n,p} = Y_n\, \xi_{n,p}^{(0)} \,, \qquad
  A_{n,q} = Y_n\, A_{n,q}^{(0)}\, Y_n^\dagger \,, \qquad 
Y_n = {\rm P}\, \exp \Big[ig\int_{-\infty}^x \d s\, n\cdot A_{us}(ns)\Big]\,,
\eeq
one can decouple at leading order in $\lambda$ the ultrasoft gluons from the
collinear Lagrangian.  Thus, nonperturbative usoft effects can be made explicit
through factors of $Y_n$ in operators.  This way SCET simplified the proofs of
classic factorization theorems (e.g., Drell-Yan, DIS, etc.~\cite{Bauer:2002nz})
and allowed new ones to be proven to all orders in $\alpha_s$ (e.g., $B\to
D^+\pi^-$~\cite{BPSdpi}, $B\to D^0\pi^0$~\cite{Mantry:2003uz}).

Going to subleading order in $\lambda$ is essential if one wants to study heavy
to light transitions.  Collinear and ultrasoft quarks cannot interact at leading
order, so a particularly important term is the mixed usoft-collinear Lagrangian,
${\cal L}_{\xi q}^{(1)}$, which is suppressed by one power of $\lambda$ and
allows to couple an usoft and a collinear quark to a collinear
gluon~\cite{Beneke:2002ph}.  We shall come back to it below.

\subsection{The semileptonic form factors and $|V_{ub}|$}
\label{sec:slff}

It has been proven using SCET that at leading order in $\lqcd/Q$ ($Q=E,m_b$), to
all orders in $\alpha_s$, the semileptonic form factors for $q^2 \ll m_B^2$ can
be written as a sum of two terms~\cite{Beneke:2000wa,Bauer:2002aj,Hill:2004if},
\beq\label{slfact}
F(Q) = C_i(Q)\, \zeta_i(Q) +
  {m_B f_B f_M \over 4E^2} \int\!\ \d z\, \d x\, \d k_+\, T(z,Q)\,
  J(z,x,k_+,Q)\, \phi_M(x)\, \phi_B(k_+) \,,
\eeq
where we omitted the $\mu$-dependences.  The two terms arise from matrix
elements of distinct time ordered products of the form, for example, $\int\!
\d^4 x\, T \big[ J^{(n)}(0)\, {\cal L}_{\xi q}^{(m)}(x) \big]$, where $J^{(n)}$
is the expansion of the current and the ${\cal L}_{\xi q}$ terms can turn the
ultrasoft spectator to a collinear quark.  In Eq.~(\ref{slfact}) the second,
factorizable (or hard scattering), term only contains ultrasoft fields in the
combination $Y^\dagger h_v^{(b)}$ and $Y^\dagger q_{us}$, and is calculable in
an expansion in $\alpha_s(\sqrt{\lqcd m_b})$.  The first, nonfactorizable (or
form-factor), term satisfies symmetry relations~\cite{Charles:1998dr}.  For any
current (Dirac structure) the nonfactorizable parts of the 3 $B\to$ pseudoscalar
and the 7 $B\to$ vector meson form factors are related to just 3 universal
functions in the heavy quark limit.

The two terms are of the same order in the $\lqcd/m_b$ power counting.  The
factorizable (2nd) term contains $\alpha_s(\sqrt{\lqcd m_b})$ explicitly, but
whether the nonfactorizable (1st) term has a similar suppression at the physical
value of $m_b$, or in the $m_b\to\infty$ limit when effects of order
$\alpha_s(m_b)$ and $\alpha_s(\sqrt{m_b\lqcd})$ are fully accounted for is an
open question.  In the applications of the three often-discussed approaches, the
assumptions for organizing the expansions and making predictions are
\beq\label{approaches}
\mbox{SCET: 1st $\sim$ 2nd}\,, \qquad
\mbox{QCDF: 2nd $\sim \alpha_s\times$(1st)}\,, \qquad
\mbox{PQCD: 1st $\ll$ 2nd}\,.
\eeq
In PQCD, the definition of the (non)factorizable terms also differs from the
above.  Clearly, what is called the leading order result and what is an
$\alpha_s$ correction differs between these approaches. While some relations
between semileptonic and nonleptonic decays can be insensitive to this, others
are not.  An important example is the value of $q^2$ where the forward-backward
asymmetry in $B\to K^*\ell^+\ell^-$ vanishes, $A_{\rm FB}(q_0^2)=0$.  While
$q_0^2$ is model independent when only the nonfactorizable part of the form
factors are considered~\cite{Burdman:1998mk}, the effect of the factorizable
term is not suppressed by $\lqcd/m_b$.  Recent calculations show that they may
in fact be sizable~\cite{Beneke:2005gs}.

The determination of $|V_{ub}|$ from $B\to\pi\ell\bar\nu$ relies on measuring
the rate and calculating the form factor $f_+(q^2)$,
\beq
{\d\Gamma(\B0bar\to\pi^+\ell\bar\nu)\over \d q^2} 
  = {G_F^2 |\vec p_\pi|^3\over 24\pi^3}\, |V_{ub}|^2\, |f_+(q^2)|^2 \,.
\eeq
Unquenched calculations of $f_+$ are only available for large $q^2$ (small
$|\vec p_\pi|$)~\cite{Okamoto:2004xg,Shigemitsu:2004ft}, but experiment loses 
statistics, since the phase space is proportional to $|\vec p_\pi|^3$. Averaging
these LQCD calculations and using data in the  $q^2>16\,\GeV^2$ region yields
$|V_{ub}| = (4.13 \pm 0.62) \times
10^{-3}$~\cite{Group(HFAG):2005rb,Okamoto:2005zg}.

Some of the current $|V_{ub}|$ determinations use model dependent
parameterizations of $f_+(q^2)$ to extend the lattice results to a larger part
of the phase space, or to combine them with QCD sum rule calculations at small
$q^2$ (which tend to give smaller values for $|V_{ub}|$~\cite{Ball:2005tb}). 
Such model dependent ingredients should be avoided; given the successes of the
CKM picture, only analyses with well-defined errors are interesting.  It has
long been known that dispersion relations and the knowledge of $f_+(q^2)$ at a
few values of $q^2$ give strong bounds on its shape~\cite{Boyd:1994tt}.  The new
LQCD results revitalized this
area~\cite{Fukunaga:2004zz,Arnesen:2005ez,Becher:2005bg}, including the
possibility of using factorization and the $B\to\pi\pi$ data to constrain
$f_+(m_\pi^2)$~\cite{Arnesen:2005ez}.  Using the lattice calculations of $f_+$,
the experimental measurements and dispersion relation to constrain $f_+(q^2)$ at
all $q^2$, yields $|V_{ub}| = (3.92 \pm 0.52) \times 10^{-3}$~\cite{iain-lp05}.

\subsubsection{$B\to \rho\gamma$ vs.\ $K^*\gamma$}

The factorization formula for the form factors in Eq.~(\ref{slfact}) also
provides the basis for addressing the corrections to unity in the $SU(3)$
breaking parameters, $\xi_\gamma^{0,\pm}$, in the ratios
\beq
{{\cal B}(B^0\to \rho^0\gamma) \over {\cal B}(B^0\to K^{*0}\gamma)}
  = \frac12\, \bigg|{V_{td}\over V_{ts}}\bigg|^2 (\xi_\gamma^0)^{-2}\,, 
  \qquad
{{\cal B}(B^\pm\to \rho^\pm\gamma) \over {\cal B}(B^\pm\to K^{*\pm}\gamma)}
  = \bigg|{V_{td}\over V_{ts}}\bigg|^2 (\xi_\gamma^\pm)^{-2}\,.
\eeq
The $\xi_\gamma$ are the analogs of $\xi$ that enters for $\Delta m_{B_d} /
\Delta m_{B_s}$.  The neutral mode gives a theoretically cleaner determination
of $|V_{td}/V_{ts}|$, because the weak annihilation contribution is absent.  (It
is suppressed by $\lqcd/m_b$, but may be numerically significant and is hard to
estimate.)  So far, there is no direct LQCD calculation of $\xi_\gamma$, and I
was glad to hear  at this meeting that this may soon become possible using a
moving NRQCD action.

Recently, BELLE observed the exclusive $B \to (\rho/\omega) \gamma$
decay~\cite{Abe:2005rj}.  Assuming isospin symmetry, the average of the BELLE
and BABAR data (without including the $\omega$) is ${\cal B}(B \to \rho
\gamma)/{\cal B}(B \to K^* \gamma) = 0.017 \pm
0.006$~\cite{Group(HFAG):2005rb}.  Using $\xi_\gamma = 1.2 \pm 0.15$ for this
average~\cite{Grinstein:2000pc} implies $|V_{td}/V_{ts}| = 0.16 \pm 0.03$.  The
smallness of ${\cal B}(B \to \rho \gamma)$, due to its non-observation at BABAR,
may be a fluctuation or indicate that $\Delta m_{B_s}$ could be not near the
experimental lower bound.  While the theoretical error of this determination of
$|V_{td}/V_{ts}|$ will not become competitive with that from $\Delta
m_{B_d}/\Delta m_{B_s}$, it provides an important test of the SM, as NP could
contribuite to these decays and $B-\Bbar$ mixing differently.

\subsection{Photon polarization in $B\to K^*\gamma$ and $X_s\gamma$}

Although the $B\to X_s\gamma$ rate is correctly predicted by the SM at the 10\%
level, the measurement sums over the rates to left- and right-handed photons,
and their ratio is also sensitive to NP.  In the SM, $b$ quarks mainly decay to
$s\gamma_L$ and $\bar b$ quarks to $\bar s\gamma_R$.  This is easy to see at a
hand-waving level, considering angular momentum conservation in the two-body
$b\to s\gamma$ decay and the fact that due to the left-handed $W$ couplings the
$s$ quark is left-handed in the $m_s \ll m_b$ limit.  It also holds to all
orders in $\alpha_s$ for the dominant operator $O_7 \sim \bar s\,
\sigma^{\mu\nu} F_{\mu\nu} (m_b P_R + m_s P_L) b = \bar s\, (m_b F_{\mu\nu}^L +
m_s F_{\mu\nu}^R) b$.  Here $F_{\mu\nu}^{L,R} = \frac12 (F_{\mu\nu} \pm i
\widetilde F_{\mu\nu})$ are the field-strength tensors for $\gamma_{L,R}$, and
$\widetilde F_{\mu\nu} = \frac12 \varepsilon_{\mu\nu\rho\lambda}
F^{\rho\lambda}$. 

The only observable measured so far that is sensitive to the photon polarization
is the time dependent $CP$ asymmetry, which is proportional to $r = A(\B0bar\to
X_{f_s}\gamma_R) / A(\B0bar\to X_{f_s}\gamma_L)$.  It has been believed that $r
\propto m_s/m_b$, and therefore the SM prediction for  $S_{K*\gamma}$ [see
Eq.~(\ref{SCdef})] is at the few percent level~\cite{Atwood:1997zr}. The world
average is $S_{K*\gamma} = -0.13 \pm 0.32$, consistent with a small value.

\begin{figure}[t]
\centerline{\includegraphics[width=.17\textwidth]{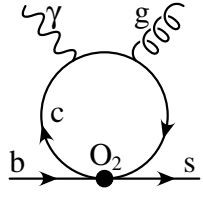}}
\caption{Dominant contribution to the "wrong-helicity" amplitude, $A(\B0bar\to
X_{f_s}\gamma_R)$.}
\label{fig:gammaR}
\end{figure}

It was recently realized that contributions from four-quark operators (see
Fig.~\ref{fig:gammaR}) give rise to $r$ not suppressed by
$m_s/m_b$~\cite{Grinstein:2004uu}.  The numerically dominant contribution is due
to the matrix element of $O_2 = (\bar c\, \gamma^\mu P_L b) (\bar s\, \gamma_\mu
P_L c)$.  Its contribution to the inclusive rate can be calculated reliably, and
at ${\cal O}(\alpha_s^2\beta_0)$ one finds $\Gamma(\B0bar\to X_s\gamma_R) /
\Gamma(\B0bar\to X_s\gamma_L) \approx 0.01$~\cite{Grinstein:2004uu}.  This
suggests that for most final states, on average, $r \sim 0.1$ should be
expected.

Experimentally most relevant is $r_{K^*}$ in the exclusive $K^*\gamma$ channel,
which can be analyzed using SCET.  (A few years ago one could have only mumbled
that the contribution of Fig.~\ref{fig:gammaR} to $r_{K^*}$ is related to higher
$K^*$ Fock states.)  In the factorization formula for the form factors,
Eq.~(\ref{slfact}), the second (factorizable) part contains an operator that
could contribute at leading order in $\lqcd/m_b$, but its $B\to K^*\gamma$
matrix element vanishes~\cite{Grinstein:2004uu}.  This proves that $r_{K^*} =
{\cal O}(\lqcd/m_b)$.  At order $\lqcd/m_b$, there are several contributions to
$A(\B0bar\to \K0bar{}^*\gamma_R)$, but there is no complete study yet.  Thus, we
can only estimate $A(\B0bar\to \K0bar{}^*\gamma_R) / A(\B0bar\to
\K0bar{}^*\gamma_L) = {\cal O}[(C_2/ 3C_7)\, (\lqcd/ m_b)] \sim 0.1$, in
qualitative agreement with the inclusive calculation.

\subsection{Comments on $B\to \tau\nu$}

\begin{figure}[t]
\centerline{\includegraphics[width=.5\textwidth]{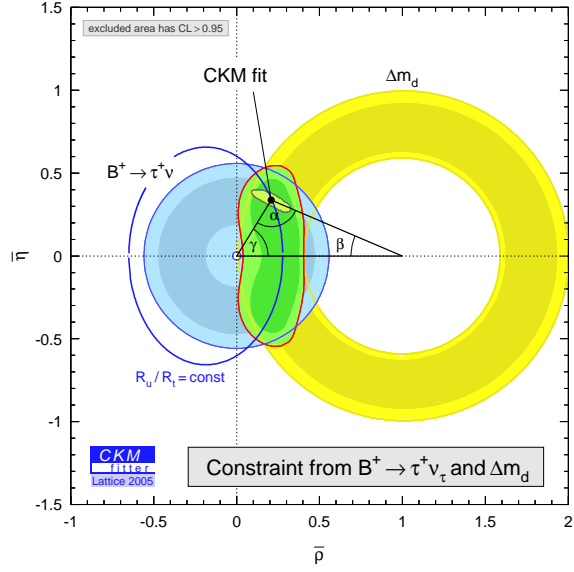}}
\caption{Constraint from $\Gamma(B\to \tau\nu)$ and $\Delta m_{B_d}$.  Shown
are the 1 and 2$\sigma$ contours.}
\label{fig:taunu}
\end{figure}

Measuring $\Delta m_{B_s}$ is not the only way to eliminate the error of $f_B$
in relating the measurement of $\Delta m_{B_d}$ to $|V_{td}|$.  The observation
of ${\cal B}(B\to \tau\nu)$ may also precede that of $\Delta m_{B_s}$.  The
$B\to \tau\nu$ measurements are usually quoted as upper bounds, but it is
already interesting to look at the data.  Figure~\ref{fig:taunu} shows the 1 and
2$\sigma$ contours with $f_B = (216 \pm 9 \pm 21)\,$MeV~\cite{Gray:2005ad}.

If the $B\to \tau\nu$ measurement was precise, $\Gamma(B\to \tau\nu) / \Delta
m_{B_d}$ would determine $|V_{ub}/V_{td}|$ independent of $f_B$ (but dependent
on $B_{B_d}$).  As shown in Fig.~\ref{fig:taunu}, we would get an ellipse in the
$\rhobar,\etabar$ plane (for fixed $V_{cb}$ and $V_{ts}$).  In the limit when
the error of $f_B$ is small, the constraints are two circles that intersect at
and angle $\alpha (\approx 100^\circ)$, which is near the right angle, providing
powerful constraints.

This is another reason why pinning down $f_B$ is very important.  While the
measurement of $B\to \tau\nu$ will improve incrementally (and will be precise
only at a super-$B$-factory), $\Delta m_{B_s}$ will almost instantly be accurate
when measured.  Measuring $\Delta m_{B_s}$ remains important not just to
determine $|V_{td}/V_{ts}|$, but to constrain NP entering $B_s$ and $B_d$ mixing
differently.  As we have emphasized, the point is to perform overconstraining
measurements and not just to determine CKM elements.

\section{Nonleptonic decays}

Having two hadrons in the final state is also a headache for us, using continuum
methods, just like it is for you, using lattice QCD.  Still, I would like to
explain some recent results for two-body nonleptonic decays.  This is the area
where the most exciting model independent results emerged recently, and it also
illustrates developments in addressing problems that will likely remain
intractable with LQCD in the near future.

\subsection{Factorization in $B\to D\pi$ type decays}

It has long been known that in $B\to M_1M_2$ decay, if the meson $M_1$ that
inherits the spectator quark from the $B$ is heavy and $M_2$ is light then
"color transparency" can justify
factorization~\cite{Bjorken:1988kk,Dugan:1990de,Politzer:1991au}. 
Traditionally, naive factorization refers to the hypothesis that one can
estimate matrix elements of the four-quark operators by grouping the quark
fields into a pair that can mediate $B\to M_1$ transition, and another pair that
describes $\mbox{vacuum} \to M_2$ transition.  We will call factorization the
systematic separation of the physics associated with different momentum scales
in a decay. For $\B0bar\to D^{(*)+}\pi^-$ these notions coincide, and amount to
showing that the contributions of gluons between the pion and the heavy meson
are either calculable perturbatively or are suppressed by $\lqcd/m_{b,c}$.  This
was proven to order $\alpha_s$~\cite{Politzer:1991au} and
$\alpha_s^2$~\cite{BBNS}, and subsequently to all orders in perturbation
theory~\cite{BPSdpi}.  Thus, up to order $\lqcd/Q$ and $\alpha_s(Q)$ corrections
($Q = E_\pi, m_{b,c}$),
\beq\label{BDpifact}
\langle D^{(*)}\pi |\, O_i(\mu_0) | B\rangle 
  = iN_{(*)}\, F_{B\to D^{(*)}}\, f_\pi \int_0^1\! \d x\;
  T(x,\mu_0,\mu)\, \phi_\pi(x,\mu) \,.
\eeq
The $O_i$ are operators in the effective Hamiltonian~\cite{Buchalla:1995vs},
$N_{(*)}$ is a normalization factor, $F_{B\to D^{(*)}}$ is the $B\to D^{(*)}$
form factor at $q^2=m_\pi^2$, $f_\pi$ is the pion decay constant, $T$ is a
perturbatively calculable function, and $\phi(x)$ is the pion wave function.

\begin{figure}[t]
\centerline{\includegraphics[width=.21\textwidth]{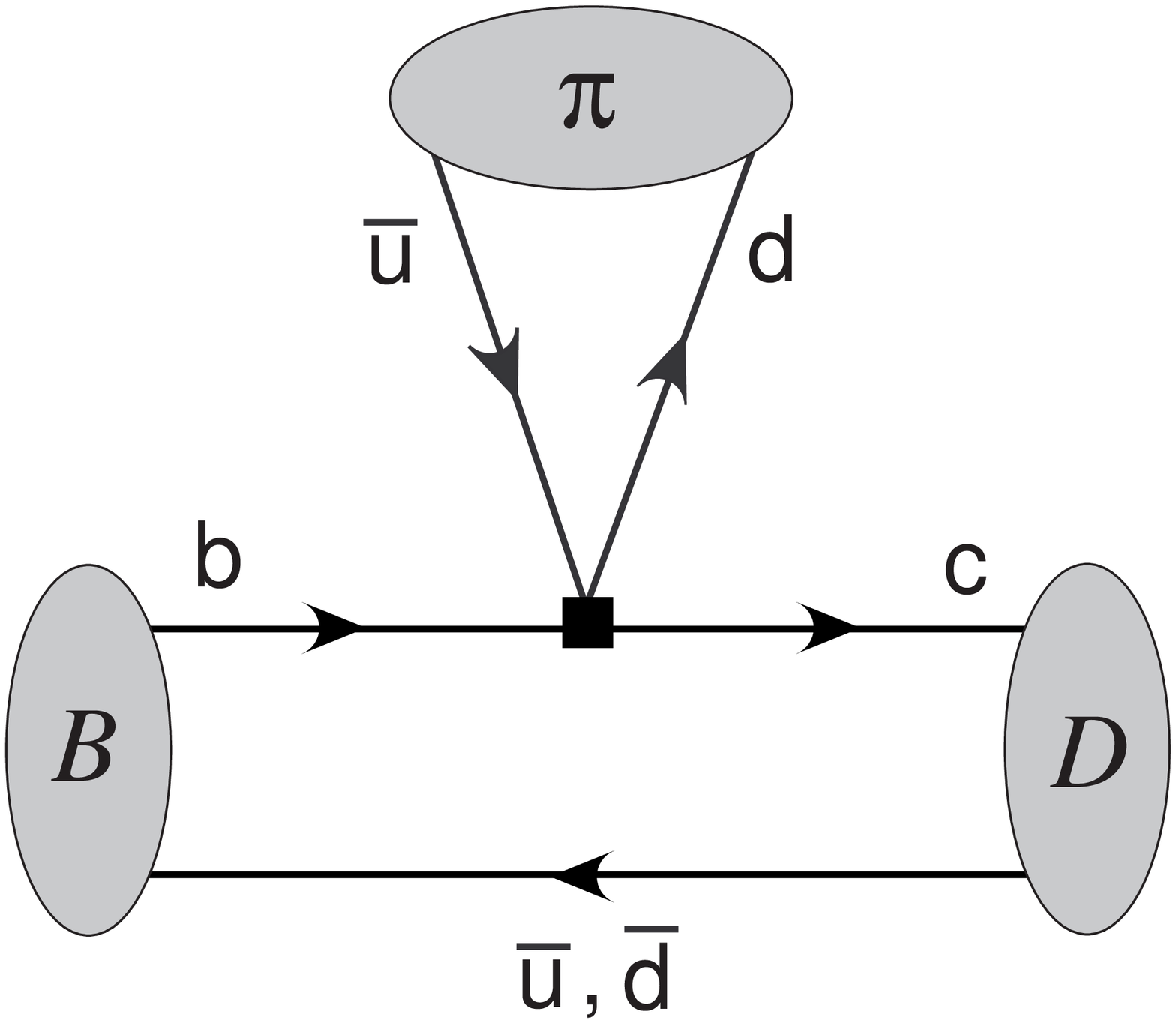} \hspace*{1cm}
  \includegraphics[width=.22\textwidth]{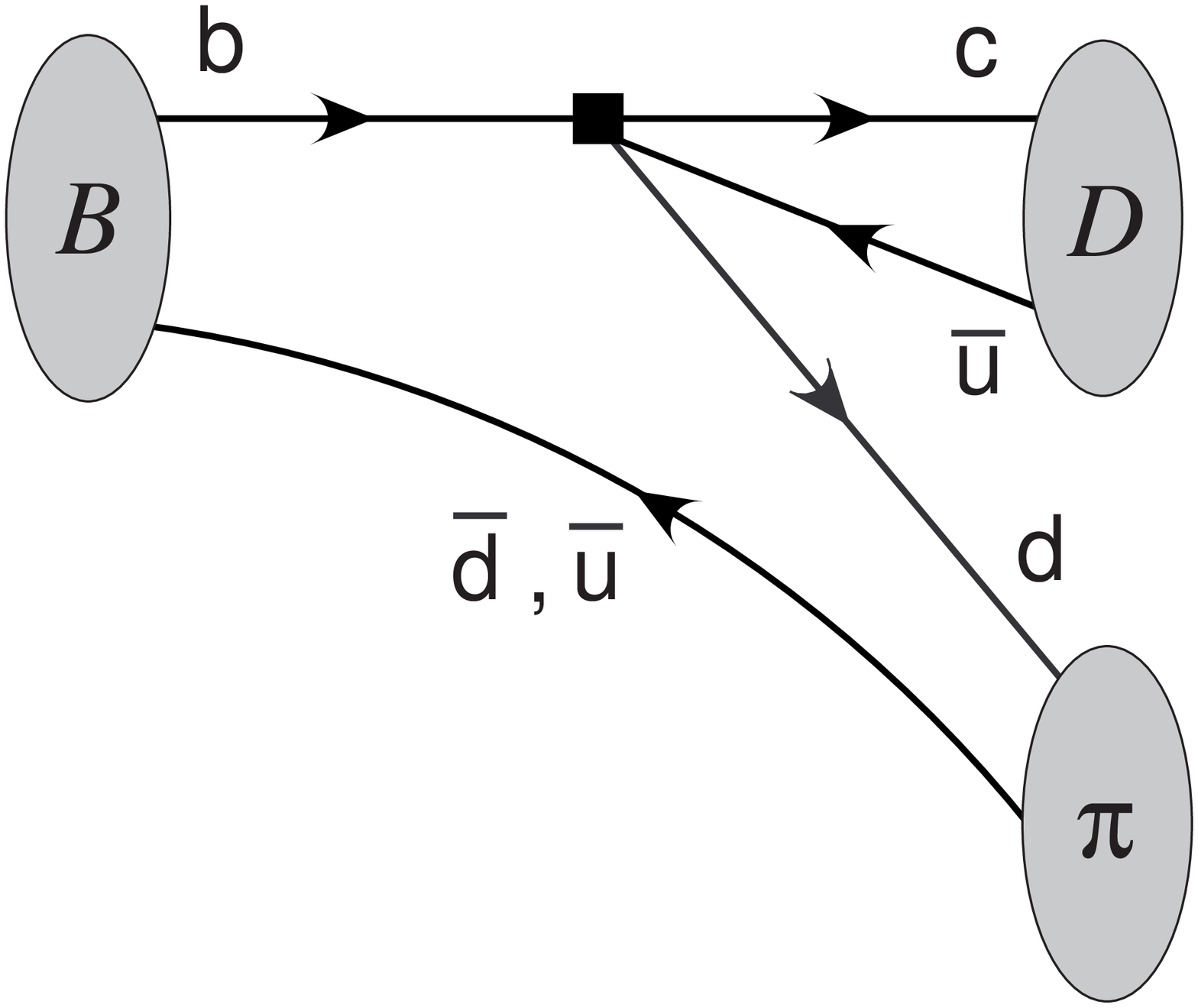} \hspace*{1cm}
  \includegraphics[width=.21\textwidth]{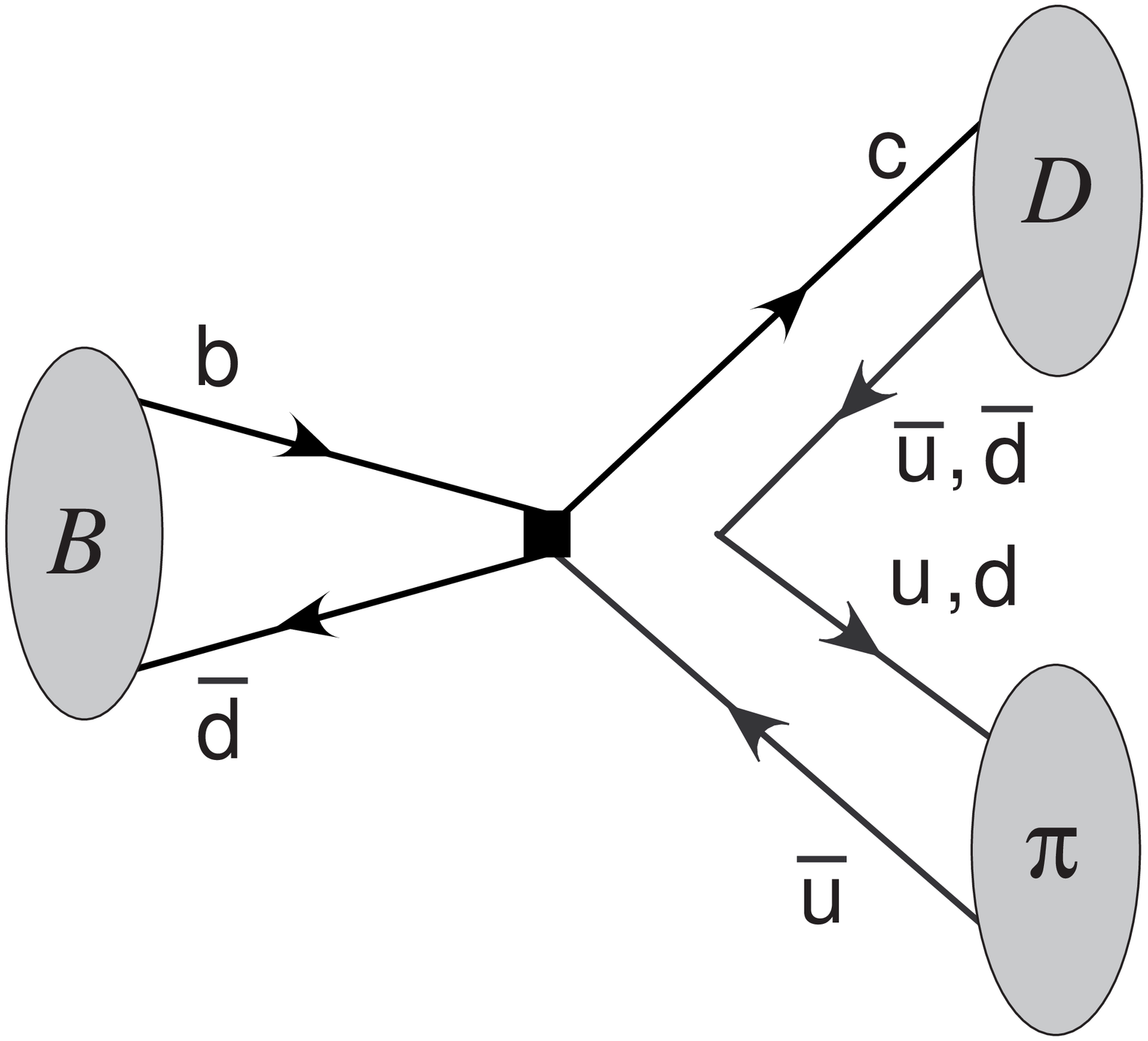}}
\vspace{-0.2cm}
\hbox{\small\hspace{2.7cm} Tree \hspace{2.7cm} Color-suppressed
  \hspace{2.2cm}  Exchange}
\caption{Diagrams for $B$ decays, giving amplitudes $T$, $C$, and $E$.  Decays
to $D^+\pi^-$, $D^0\pi^-$, $D^0\pi^0$ receive $T$ and $E$, $T$ and $C$, and $C$
and $E$ contributions, respectively (from~\cite{Mantry:2003uz}).}
\label{fig:BDpi}
\end{figure}

There are three contributions to the $B\to D\pi$ amplitudes, shown in
Fig.~\ref{fig:BDpi}.  SCET implies the power counting $T = {\cal O}(1)$ and $C,E
= {\cal O}(\lqcd/Q)$.  In decays such as $\B0bar\to D^+\pi^-$, which have $T$
and $C$ contributions, factorization has been observed to work at the 5--10\%
level.  For these rates naive factorization also holds in the large $N_c$ limit
(up to $1/N_c^2$ corrections), so detailed tests are needed to establish the
mechanism responsible for factorization.  At the current level of accuracy,
there is no evidence for factorization becoming a worse approximation as the
invariant mass of the "light" final state increases~\cite{Ligeti:2001dk}, which
is expected at some level if the heavy quark limit is important.  The heavy
quark limit also implies ${\cal B}(B^-\to D^{(*)0}\pi^-) / {\cal B}(\Bbar^0\to
D^{(*)+}\pi^-) = 1 + {\cal O}(\lqcd/Q)$, but experimentally this ratio is around
$1.8\pm0.2$ (for all four combinations of $D,D^*$ and $\pi,\rho$ final states),
indicating ${\cal O}(30\%)$ power corrections.

The $B_s\to D_s\pi$ decay only proceeds via a $T$ contribution, so it can help
to determine the relative size of $E$ vs.\ $C$.  CDF measured ${\cal B}(B_s\to
D_s^-\pi^+)/ {\cal B}(B^0\to D^-\pi^+) \simeq 1.35 \pm 0.43$~\cite{cdfds} (using
the production ratio $f_s/f_d = 0.26\pm 0.03$), the central value of which
suggests that $C$ and $E$ may be comparable~\cite{llsw}.  Since factorization
relates the tree amplitudes to the semileptonic form factors, LQCD could play an
important role by computing the $SU(3)$ breaking in the $B_s\to D_s\ell\bar\nu$
vs.\  $B\to D\ell\bar\nu$ form factors.  This is a "gold-plated" quantity, which
I hope may be found on some people's computers in the audience.

\subsubsection{Color suppressed $B\to D^{(*)0} M^0$ decays}

The $B^0\to D^0\pi^0$ decay receives only $C$ and $E$ contributions, which are
suppressed by $\lqcd/Q$.  These rates were beleived to be untractable until it
was observed that a single class of power suppressed SCET$_{\rm I}$ operators
give rise to these decays~\cite{Mantry:2003uz}.  To turn the ultrasoft spectator
quark in the initial $B$ into a collinear quark in the outgoing $\pi^0$, time
ordered products with two factors of ${\cal L}_{\xi q}^{(1)}$ are needed.  A
factorization formula that separates the different scales was
proven~\cite{Mantry:2003uz}
\beq\label{colorsup}
A(D^{(*)0}M^0) = N_0^M\! \int\!\! \d z\, \d x\, \d k_1^+ \d k_2^+\, 
  T^{(i)}(z)\, J^{(i)}(z,x,k_1^+,k_2^+)\, 
  S^{(i)}(k_1^+, k_2^+) \phi_M(x) + \ldots \,,
\eeq
where $i=0,8$ label singlet and octet color structures and, for example,
\beq
S^{(0)}(k^+_1,k^+_2) = 
  \frac{\langle D^{0}(v') | (\bar h_{v'}^{(c)}S)\, \nslash P_L
  (S^\dagger h_v^{(b)}) (\bar d S)_{k^+_1}\, \nslash P_L (S^\dagger u)_{k^+_2} 
  | \bar B^0(v)\rangle}{\sqrt{m_B m_D}} \,,
\eeq
where the $S$ is a soft Wilson line in SCET$_{\rm II}$. This is quite a
different factorization than Eq.~(\ref{BDpifact}), as $S^{(i)}(k^+_1,k^+_2)$
which contain the low energy nonperturbative physics is the matrix element of a
four-quark operator.  It depends on the direction of the outgoing pion, $n$, and
is a complex quantity, indicating that factorization can accommodate a
nonperturbative strong phase.

\begin{figure}[t]
\centerline{\includegraphics*[width=.5\textwidth]{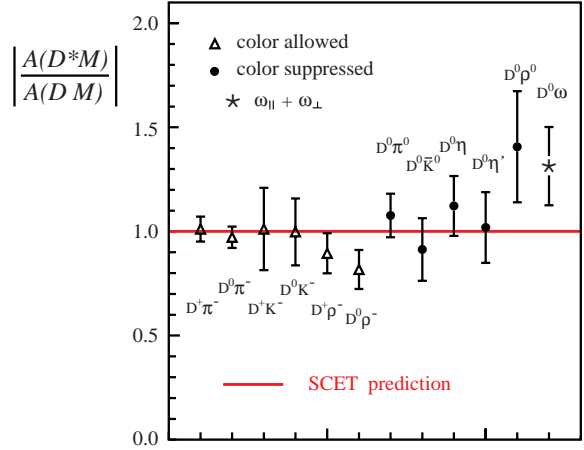}}
\caption{Ratios of the amplitudes $A(B\to D^{*0} M^0) / A(B\to D^0 M^0)$
extracted from data (from~\cite{Blechman:2004vc}).}
\label{fig:mantry}
\end{figure}

Still, these formulae allow several nontrivial predictions to be made.  The
separation of scales allows one to use HQS for $S^{(i)}(k^+_1,k^+_2)$ without
encountering $E_\pi/m_c = {\cal O}(1)$ corrections, which would occur if one
attempted to use HQS for this decay in full QCD.  At leading order one finds the
predictions $A(B\to D^{*0} M^0) / A(B\to D^0 M^0) =
1$~\cite{Mantry:2003uz,Blechman:2004vc}, similar to final states with charged
mesons.  These are compared with the data in Fig.~\ref{fig:mantry}, where the
${\scriptstyle \triangle} = 1$ relations follow from naive factorization and
HQS, however, the $\bullet = 1$ relations for the neutral modes do not, and
constitute a profound prediction of SCET not foreseen by model calculations.
SCET also predicted equal strong phases between the $I = 1/2$ and $3/2$
amplitudes in $B\to D\pi$ and $D^*\pi$.  The measurements, made after the
prediction, give $\delta(D\pi) = (28 \pm 3)^\circ$ and $\delta(D^*\pi) = (32 \pm
5)^\circ$~\cite{Pirjol:2004yf}.

\subsubsection{$\Lambda_b \to \Lambda_c\pi$ and $\Sigma_c^{(*)}\pi$ decays}

\begin{figure}[t]
\centerline{\includegraphics[width=.23\textwidth]{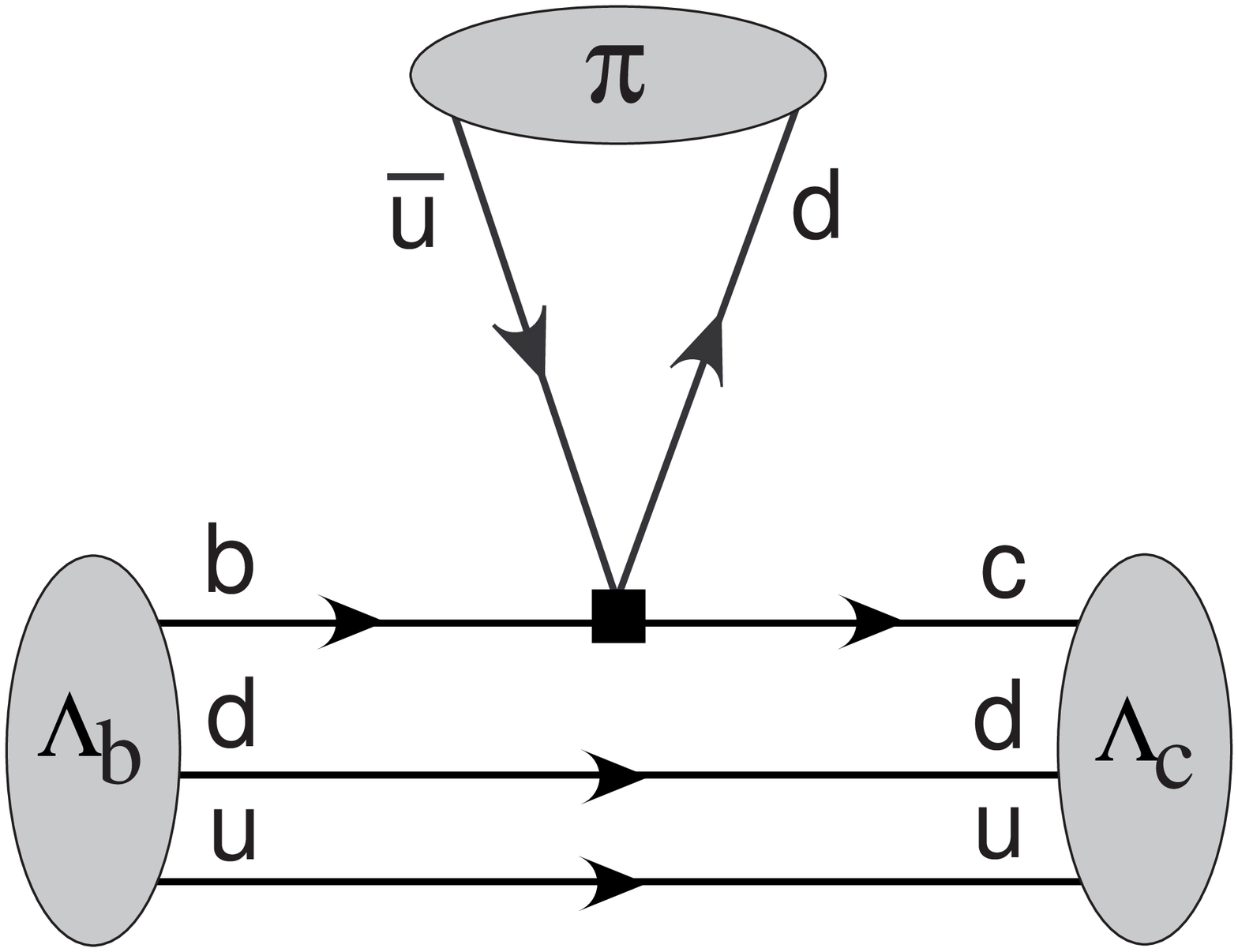} \hspace*{0.1cm}
  \includegraphics[width=.23\textwidth]{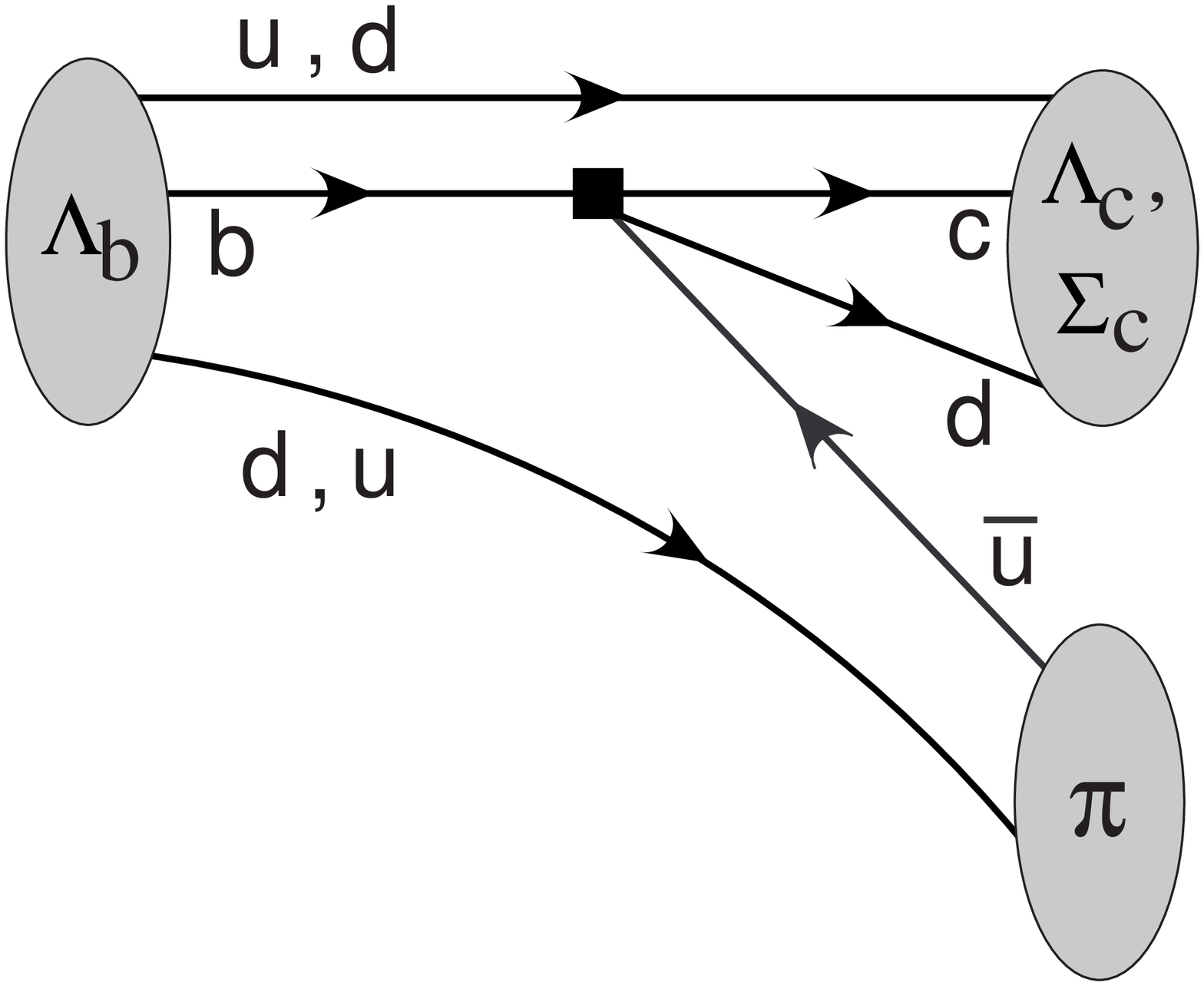} \hspace*{0.1cm}
  \includegraphics[width=.22\textwidth]{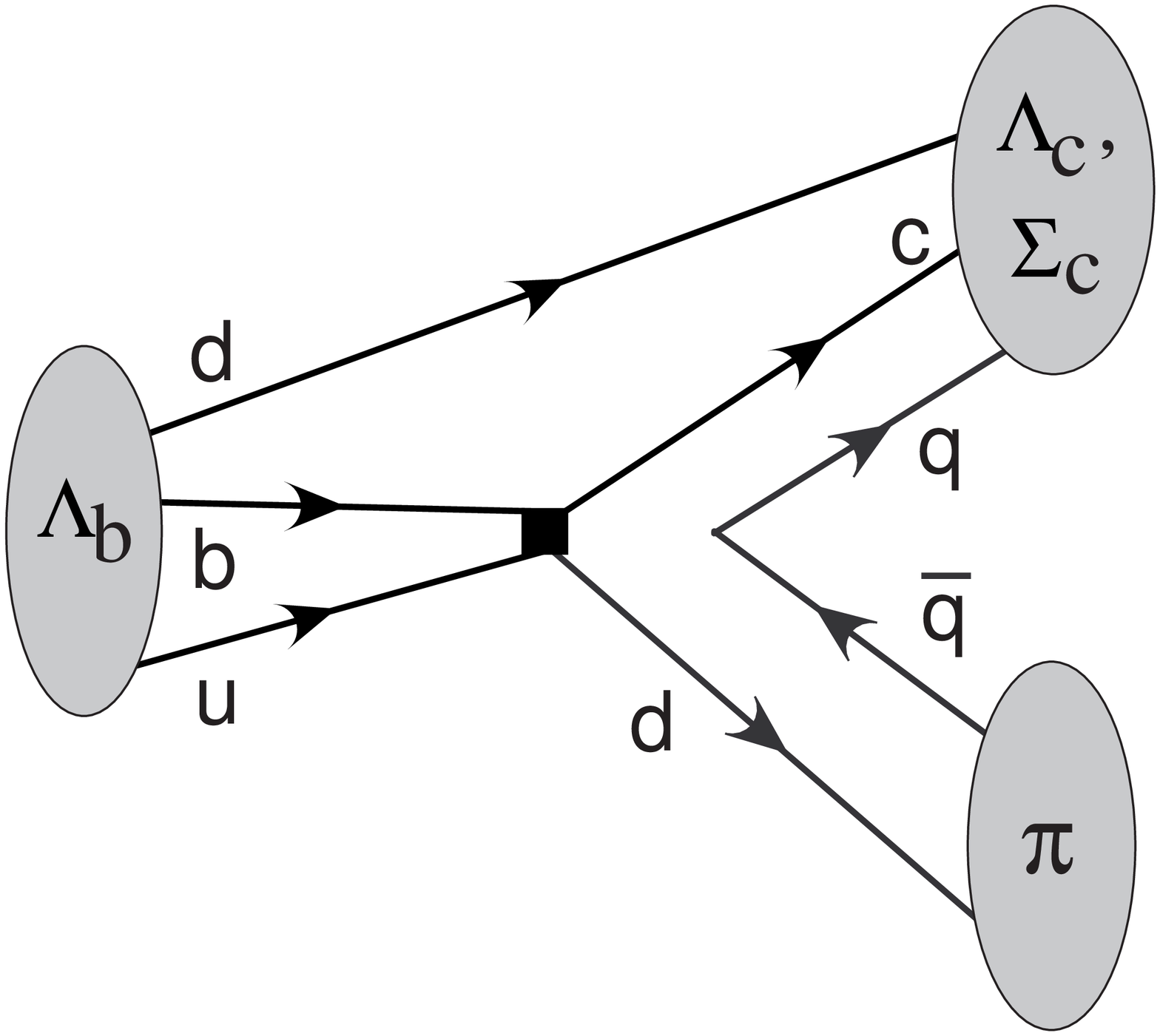} \hspace*{0.1cm}
  \includegraphics[width=.23\textwidth]{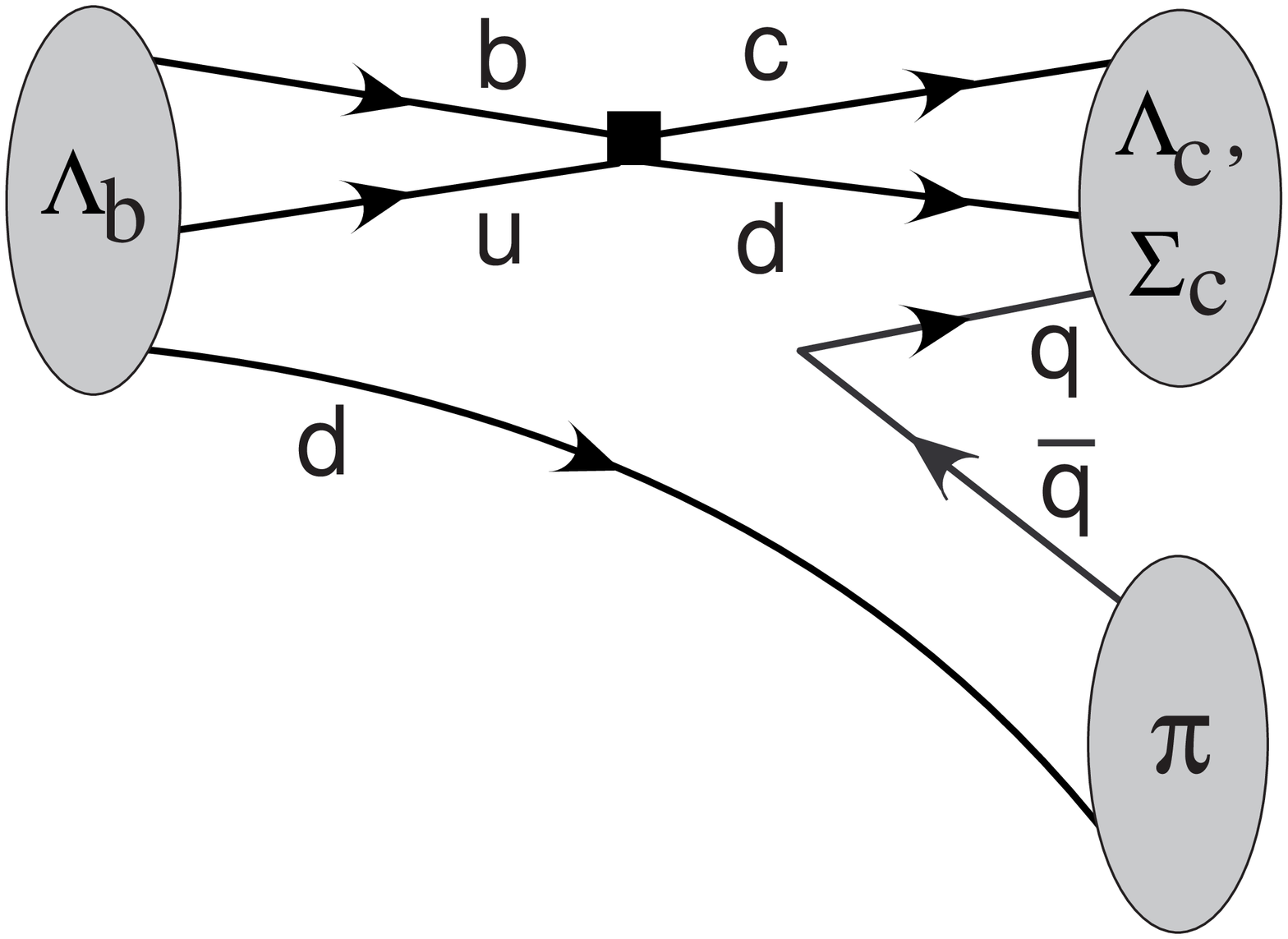}}
\vspace{-0.2cm}
\hbox{\small\hspace{1.5cm} Tree \hspace{1.9cm} Color-commensurate
  \hspace{1.2cm}  Exchange \hspace{2.25cm} Bow tie}
\caption{Diagrams for $\Lambda_b$ decays, giving amplitudes $T$, $C$, $E$, and
$B$. Decay to $\Lambda_c$ gets contributions from all four terms.  Decays to
$\Sigma_c^{(*)}$ [$\Xi_c$] do not have $T$ [$T$ and $C$] contributions
(from~\cite{llsw}).}
\label{fig:diagrams}
\end{figure}

Factorization for baryon decays does not follow from large $N_c$, but it still
holds in the heavy quark limit at leading order in $\lqcd/Q$, providing an
interesting test.  There are four contributions to $\Lambda_b\to \Lambda_c^+
\pi^-$, as shown in Fig.~\ref{fig:diagrams}, and SCET implies the power counting
$T = {\cal O}(1)$, $C,E = {\cal O}(\lqcd/Q)$, and $B = {\cal
O}(\lqcd^2/Q^2)$~\cite{llsw}.  The usual factorization relation connects ${\cal
B}(\Lambda_b\to \Lambda_c \pi)$ measured by CDF~\cite{cdflambdab} to
$\d\Gamma(\Lambda_b\to \Lambda_c \ell\bar\nu)/\d q^2$ at $q^2=m_\pi^2$ (maximal
recoil).  Thus, either an experimental measurement or a LQCD calculation of the
$\Lambda_b\to \Lambda_c \ell\bar\nu$ Isgur-Wise function would allow this
relation to be tested.

For the power suppressed $\Lambda_b\to \Sigma_c^{(*)} \pi$ decays (we denote
$\Sigma_c = \Sigma_c(2455)$ and $\Sigma_c^*=\Sigma_c(2520)$), naive
factorization makes no sense, since the $\Lambda_b \to \Sigma_c^{(*)}$ form
factors violate isospin, whereas $\Lambda_b\to \Sigma_c^{(*)} \pi$ can occur at
a power suppressed level (compared to $\Lambda_b\to \Lambda_c \pi$) without
isospin violation.  An analysis similar to meson decays yields~\cite{llsw}
\beq
{\Gamma(\Lambda_b\to\Sigma_c^*\pi) \over
  \Gamma(\Lambda_b\to\Sigma_c\pi)}
= {\Gamma(\Lambda_b\to\Sigma_c^{*0}\rho^0) 
  \over \Gamma(\Lambda_b\to\Sigma_c^0\rho^0)}
= 2 + {\cal O} \big[\lqcd/Q\,,\, \alpha_s(Q) \big] \,.
\eeq
Interestingly, this ratio is predicted to be twice as large as the similar
ratios in the meson sector discussed
above~\cite{Mantry:2003uz,Mantry:2004pg,Blechman:2004vc}.  Isospin symmetry
implies $\Gamma(\Lambda_b \to \Sigma_c^{(*)0} \pi^0) = \Gamma(\Lambda_b \to
\Sigma_c^{(*)+} \pi^-)$, and similarly for the $\rho$'s.  The second ratio is
useful because it has no $\pi^0$'s in the final state, and therefore it may be
easier to measure at a hadron collider (in the first ratio a $\pi^0$ is
unavoidable either from $\Lambda_b \to \Sigma_c^{(*)0} \pi^0$ or from $\Lambda_b
\to \Sigma_c^{(*)+} \pi^- \to \Lambda_c^+ \pi^0\pi^-$).

\subsection{Charmless $B\to M_1M_2$ decays}

Factorization is more complicated for charmless $B$ decays, but I want to talk
about some aspects, because these processes are in principle sensitive to new
physics.  In this case there is limited consensus about the implications of the
heavy quark limit.  In SCET a factorization formula has been 
proven~\cite{Bauer:2004tj,Chay:2003ju,bbnslight}
\beqa\label{pipifact}
A(B\to M_1 M_2) = A_{c\bar c} 
&+& N \Big[f_{M_2}\, \zeta^{BM_1} \!\int\! \d u\,
  T_{2\zeta}(u)\, \phi_{M_2}(u) \nn\\*
&&{} + f_{M_2} \!\int\! \d z\, \d u\, T_{2J}(u,z)\, 
  \zeta_J^{BM_1}(z)\, \phi_{M_2}(u) + (1 \leftrightarrow 2) \Big] .
\eeqa
Here $\zeta_J^{BM_1}(z) = f_{M_1} f_B \int\! \d x\, \d k_+\, J(z,x,k_+)\,
\phi_{M_1}(x)\, \phi_B(k_+)$ is the same object that appears in the $B\to M_1$
form factors in Eq.~(\ref{slfact}).  Therefore, the relations to semileptonic
decays do not require an expansion in $\alpha_s(\sqrt{\lqcd Q})$.  As we saw for
the semileptonic form factors in Sec.~\ref{sec:slff}, the nonfactorizable (1st)
and factorizable (2nd) terms in square brackets are of the same order in
$\lqcd/Q$.  Similar to Eq.~(\ref{approaches}), the different ways to make
quantitative predictions, usually labelled
SCET~\cite{Bauer:2004tj,Bauer:2004dg,Bauer:2005kd},
QCDF~\cite{bbnslight,Beneke:2003zv}, and PQCD~\cite{keumetal} treat the two
terms differently, which has important implications for the predictions and fits
to the data.  The $T$'s are always calculated perturbatively.  In SCET, one fits
both the $\zeta$'s and $\zeta_J$'s; in QCDF, one fits the $\zeta$'s and
calculates the $\zeta_J$'s perturbatively; and in PQCD the factorizable (2nd)
terms dominate and depend on $k_\perp$.

The $A_{c\bar c}$ term is a possible nonperturbative contribution due to charm
loops, the power counting for which is subject to debate~\cite{pcdedbate}.  A
large $A_{c\bar c}$ amplitude was found in the SCET fit to
$B\to\pi\pi$~\cite{Bauer:2004tj}, or adding a free parameter to the leading
order QCDF result~\cite{charmloops}.  There are several model dependent
calculations of this effect, referred to as "long distance charm loops",
"charming penguins", and "$D\Dbar$ rescattering", all of which is the same
unknown physics that may yield strong phases, transverse polarization, and other
"surprises".  If one views $A_{c\bar c}$ as a nonperturbative term that has to
be fit from data, then it can accommodate the sizable strong phase observed in
$A_{K^-\pi^+}$ [see Eq.~(\ref{KpidirectCP}) where the ratio of the two
interfereng amplitudes is known to be not near unity], which is hard to
reproduce in QCDF (and is in the ballpark of earlier PQCD predictions).  A
fairly generic feature of QCDF is that it tends to predict small direct $CP$
asymmetries due to the $\alpha_s$ suppression of the factorizable contributions.

Another area where these effects may be important is the longitudinal
polarization fraction in charmless $B$ decays to two vector mesons, such as
$B\to \phi K^*$, $\rho\rho$, and $\rho K^*$.  It was argued~\cite{Kagan:2004uw}
that the chiral structure of the SM and the heavy quark limit imply that these
decays must have longitudinal polarization fractions near unity, $1 - f_L =
{\cal O}(1/m_b^2)$.  It is now well-established that $f_L(\phi K^*) \approx
0.5$, while $f_L(\rho\rho)$ is near unity.  Several explanations have been
proposed why the data may be consistent with the
SM~\cite{Bauer:2004tj,Kagan:2004uw,Colangelo:2004rd,Hou:2004vj}.  While
$f_L(\phi K^*)$ may be a result of new physics contributions (just like
$A_{K^-\pi^+}$), we cannot rule out at present that it is simply due to SM
physics.

\subsubsection{$B\to \pi\pi$ amplitudes and $CP$ asymmetries}
\label{sec:pipi}

Since the error of $\alpha$ from the $B\to \pi\pi$ isospin analysis is large
(currently $|\Delta\alpha| < 37^\circ$, see Sec.~\ref{sec:alpha}), it will take
a long time for this measurement to become precise.  This makes it interesting
to use more theoretical input to determine $\alpha$ without the least precisely
known ingredient of the isospin analysis, the direct $CP$ asymmetry in $B\to
\pi^0\pi^0$ in Eq.~(\ref{C00}), $C_{00}$.  The $\B0bar$ and $B^-$ amplitudes to
the three possible $\pi\pi$ final states are
\beqa\label{pipiampl}
\ov A_{+-} &=& - \lambda_u (T+P_{u}) - \lambda_c P_c - \lambda_t P_{t}
  = e^{-i\gamma}\, T_{\pi\pi} - P_{\pi\pi} \,, \nn\\
\sqrt2\, \ov A_{00} &=& \lambda_u (-C+P_u) + \lambda_c P_c + \lambda_t P_t
  = e^{-i\gamma}\, C_{\pi\pi} + P_{\pi\pi} \,, \nn\\
\sqrt2\, \ov A_{-0} &=& -\lambda_u(T+C) 
  = e^{-i\gamma} (T_{\pi\pi} + C_{\pi\pi}) \,.
\eeqa
Factorization predicts ${\rm arg}(T/C)= {\cal O}(\alpha_s,\Lambda/m_b)$, which
could eliminate the need for $C_{00}$~\cite{Bauer:2004dg}.  Due to the unknown
size of $A_{c\bar c}$, however, the implications for the physically observable
amplitudes $T_{\pi\pi}$ and $C_{\pi\pi}$ are less clear, because they are
combinations of trees and penguins.

In SCET, $P_c$ is treated as ${\cal O}(1)$, and therefore the $P_t$ term is
eliminated using $\lambda_u + \lambda_c + \lambda_t = 0$.  Then the $P_{\pi\pi}$
term has no weak phase, as shown in Eq.~(\ref{pipiampl}).  In QCDF, $P_t$ is
observed to contain "chirally enhanced" corrections (a misnomer for terms
proportional to $m_\pi^2/(m_u m_b) \sim \lqcd/m_b$) while $P_c$ is argued to be
small, and therefore the $P_c$ term is eliminated using unitarity.  Then what is
meant by $T_{\pi\pi}$ and $P_{\pi\pi}$ changes, and the $P_{\pi\pi}$ term has a
weak phase $e^{i\beta}$.  Both approaches agree that $P_u$ is calculable and
small.

\begin{figure}[t]
\centerline{\includegraphics[width=.5\textwidth]{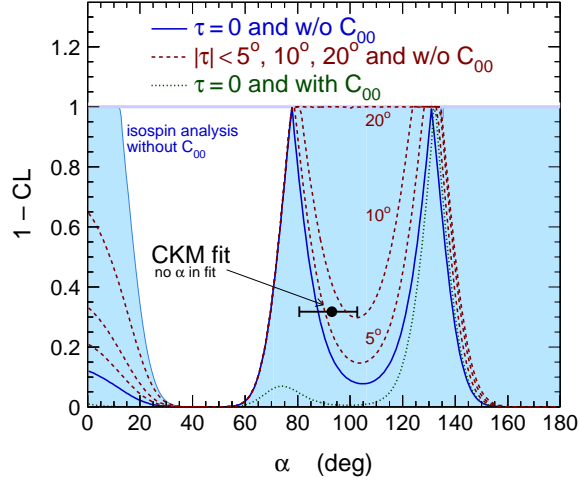}}
\caption{CL of $\alpha$ imposing $\tau = 0$ with (solid) and without (dotted)
the $C_{00}$ data.  The constraints imposing $|\tau| <5^\circ, 10^\circ,
20^\circ$ are also shown (from~\cite{Grossman:2005jb}). The shaded region is the
same as the green region in~Fig.~\protect\ref{fig:alphapipi}.}
\label{fig:pipi}
\end{figure}

The central values of the $B\to\pi\pi$ data suggest significant corrections to
the terms included in either approaches~\cite{Grossman:2005jb,Feldmann:2004mg}.
Factorization predicts that the strong phase between the "tree" amplitudes in
$\pi^+\pi^-$ and $\pi^-\pi^0$ is small, $\tau \equiv {\rm
arg}[T_{\pi\pi}/(C_{\pi\pi}+T_{\pi\pi})] = {\cal O}(\alpha_s,\Lambda/m_b)$.  As
shown in Fig.~\ref{fig:pipi}, imposing $\tau = 0$, the present data yields
$\alpha \approx 78^\circ$ (without using $C_{00}$), somewhat below the
measurement in Eq.~(\ref{alpha}).  Conversely, one can ask what the SM CKM fit
implies for $\tau$.  The result is $\tau \sim 30^\circ$~\cite{Grossman:2005jb},
both in the convention of Eq.~(\ref{pipiampl}) and the one used in QCDF.  There
are several possibile resolutions: (i) $2\sigma$ level fluctuations in the data;
(ii) large power corrections to $T$ or\,/\,and~$C$;\break (iii) large up
penguins; (iv) large weak annihilation; (v) or maybe something beyond the SM. 
It will be fascinating to find out which of these is the right explanation.

In QCDF it is problematic to accommodate the large $B\to\pi^0\pi^0$ rate in
Eq.~(\ref{pi0pi0})~\cite{bbnslight}.  This by itself is not an issue in SCET,
since the factorizable ($\zeta_J$) term in Eq.~(\ref{pipifact}) that also
determines the $B\to \pi^0\pi^0$ rate is fitted from the
data~\cite{Bauer:2004tj}.  Color suppression is ineffective because $1/N_c$ is
multiplied by the inverse moment of the $\pi$ distribution amplitude, $\langle
\bar u^{-1}\rangle_\pi = \int_0^1 \d u\, \phi_\pi(u)/(1-u)$, which is around 3. 
The $\zeta_J$ terms also depend on the similar inverse moment of the $B$
light-cone distribution amplitude, $\langle k_+^{-1}\rangle_B = \int\! \d k_+\,
\phi_B(k_+)/k_+$, and the SCET fit favors $\langle k_+^{-1}\rangle_B \sim
1/(100\,\MeV)$~\cite{Pirjol:2005ps} significantly above most QCD sum rule
calculations, which give $\sim\!1/(450\,\MeV)$~\cite{Braun:2003wx}.

While in $B\to\pi\pi$ the complications are due to the interference of
comparable $b\to u$ tree and $b\to d$ penguin processes, $B\to K\pi$ decays are
sensitive to the interference of $b\to u$ tree and the dominant $b\to s$ penguin
contributions.  The challenge is if one can make sufficiently precise SM
predictions to be sensitive to NP.  Besides the precise measurement of
$A_{K^-\pi^+}$ in Eq.~(\ref{KpidirectCP}), another interesting feature of the
data is the almost $4\sigma$ difference, $A_{K^-\pi^0} - A_{K^-\pi^+} = 0.15 \pm
0.04$.  This appears to be at odds with factorization (unless subleading terms
are very important, in which case the $\lqcd/m_b$ expansion itself becomes
questionable), because the strong phase between the penguin and the
color-allowed tree amplitude is predicted to be the same as the phase between
the penguin and the color-suppressed tree (up to $\lqcd/Q$ corrections).  A
possible resolution is a large enhancement of electroweak penguins, or NP with
the same flavor structure~\cite{Buras:2003yc}.  

These are fascinating developments, however, more work and data are needed to
understand why some predictions work better than $10\%$, while others receive
${\cal O}(30\%)$ corrections.  Hopefully, the role of charming penguins,
chirally enhanced terms, annihilation contributions, etc., can be clarified
soon.  We now have the tools to try to address these questions.

\OMIT{
\begin{table}[t]
\centerline{\begin{tabular}{|l|ll|}  
\hline
\multicolumn{1}{|c|}{Decay mode}  &  \multicolumn{1}{c}{${\cal B}\ [10^{-6}]$}
  &  \multicolumn{1}{c|}{$A_{CP}$}
\\ \hline\hline
$\B0bar\to \pi^+K^-$  &  $18.9 \pm 0.7$  &  $-0.12 \pm 0.02$ \\
$B^-\to \pi^0K^-$  &  $12.1 \pm 0.8$  &  $+0.04 \pm 0.04$ \\
$B^-\to \pi^-\K0bar$  &  $24.1 \pm 1.3$  &  $-0.02 \pm 0.04$ \\
$\B0bar\to \pi^0\K0bar$  &  $11.5 \pm 1.0$  &  $-0.02 \pm 0.13$ \\ \hline
\end{tabular}}
\caption{World average $CP$-averaged $B\to\pi K$ branching ratios, and $CP$
asymmetries.}
\label{tab:Kpi}
\end{table}
}

\section{Outlook and Conclusions}
\label{sec:fin}

The $B$ factories have provided a spectacular confirmation of the CKM picture. 
More interesting than the actual determinations of CKM elements is that
overconstraining measurements tested the CKM picture, and we can even bound
flavor models with more parameters than the SM.  In particular, the comparison
between tree- and loop-level measurements severely constrain NP in $B-\Bbar$
mixing.  For this, the lattice results on the decay constants and bag parameters
are crucial; without it we would not yet be able to really constrain these
models.  This illustrates again that the program as a whole is a lot more
interesting than any single measurement, since it is the multitude of
overconstraining measurements and their correlations that carries the most
interesting information.

Having seen these impressive measurements, one may ask where we go from here in
flavor physics?  Whether we see signals of flavor physics beyond the SM will be
decisive.  The existing measurements could have shown deviations from the SM,
and if there are new particles at the TeV scale, new flavor physics could show
up "any time".  If NP is seen in flavor physics then we will want to study it in
as many different processes as possible.  If NP is not seen in flavor physics,
then it is interesting to achieve what is theoretically possible, thereby
testing the SM at a much more precise level.  Even in the latter case, flavor
physics will give powerful constraints on model building in the LHC era, once
the masses of some new particles are known.

The present status of some of the cleanest measurements and my estimates of the
theoretical limitations (using continuum methods) are summarized in
Table~\ref{tab:future}.  The sensitivity to NP will not be limited by hadronic
physics in many measurements for a long time to come.

\begin{table}[t]
\centerline{
\begin{tabular}{|l|c|c|}
\hline
Measurement (in SM)  &  Theoretical limit  &  Present error  \\ 
\hline\hline 
$B\to \psi K$ \ ($\beta$)  &  $\sim 0.2^\circ$  &  $1.3^\circ$ \\
$B\to \eta' K,\ \phi K,$ ($\beta$)  
  &  $\sim 2^\circ$  &  $5,\ 10^\circ$\\
$B\to \rho\rho,\ \pi\pi,\ \rho\pi$ \ ($\alpha$)  
  &  $\sim 1^\circ$  &  $\sim13^\circ$\\
$B\to DK$ \ ($\gamma$)  &  $\ll 1^\circ$  &  $\sim 20^\circ$  \\
$B_s\to \psi\phi$ \ ($\beta_s$)  &  $\sim 0.2^\circ$  &  --- \\ 
$B_s\to D_sK$ \ ($\gamma-2\beta_s$)  &  $\ll 1^\circ$  &  --- \\
\hline
$|V_{cb}|$  &  $\sim 1\%$  &  $\sim2\%$ \\
$|V_{ub}|$  &  $\sim 5\%$  &  $\sim10\%$ \\
$B\to X_s \gamma$  &  $\sim 5\%$  &  $\sim10\%$ \\
$B\to X_s \ell^+\ell^-$  &  $\sim 5\%$  &  $\sim20\%$ \\
$B\to X_s\nu\bar\nu, K^{(*)} \nu\bar\nu$  &  $\sim 5\%$ &  ---  \\
\hline
$K^+\to \pi^+\nu\bar\nu$  &  $\sim 5\%$  &  $\sim70\%$ \\
$K_L\to \pi^0\nu\bar\nu$  &  $< 1\%$  &  --- \\
\hline
\end{tabular}}
\caption{Some interesting measurement that are far from being theory limited. 
The errors for the $CP$ asymmetries in the first box refer to the angles in
parenthesis, assuming typical values for other parameters.}
\label{tab:future}
\end{table}

\subsection{Where can lattice contribute the most?}

\begin{itemize}\vspace*{-0pt}\itemsep -2pt

\item Reducing the error of the decay constants and bag parameters remains very
important.

\item The determinations of semileptonic form factors is in the hands of LQCD. 
Besides those directly relevant for the extraction of CKM elements, the
computations of several others would also have important implications: we saw
examples for $B_s\to D_s\ell\bar\nu$, $\Lambda_b\to \Lambda_c\ell\bar\nu$, etc.

\item In addition to the $B,D\to\pi,K$ form factors, try to include the $\rho$
and $K^*$ final states (I know, the widths...), and attempt direct calculations
at larger recoil (maybe with moving NRQCD).

\item Dedicated and precise calculations of $SU(3)$ breaking in form factors and
in distribution functions could also play very important roles.

\item The light cone distribution functions of heavy and light mesons are
important for understanding nonleptonic decays, and so far most calculations use
QCD sum rules and other models.

\item More remote but worthwhile goals include the calculations of nonlocal
matrix elements, such as the inverse moment of the $B$ light-cone distribution
amplitude, $\langle k_+^{-1}\rangle_B = \int\! \d k_+\, \phi_B(k_+)/k_+$,
discussed above.  Not to mention nonleptonic decays...

\end{itemize}\vspace*{-8pt}

\subsection{Past and near future lessons}
\label{sec:conc}

The large number of impressive new results speak for themselves, so it is easy
to summarize the main lessons we have learned:

\begin{itemize}\vspace*{-8pt}\itemsep -2pt

\item $\sin2\beta = 0.687 \pm 0.032$ implies that the $B$ and $K$ constraints
are consistent, and the KM phase is the dominant source of CPV in
flavor-changing processes.

\item $S_{\psi K} - S_{\eta' K} = 0.21 \pm 0.10$ and  $S_{\psi K} - S_{\phi K} =
0.22 \pm 0.19$ are not conclusive yet, but the present central values with
$5\sigma$ significance could still signal NP.

\item First measurements of $\alpha = \big(99^{+13}_{-8}\big)^\circ$ and 
$\gamma = \big(63^{+15}_{-12}\big)^\circ$ start to give the tightest constraints
on $\rhobar,\etabar$ and the first serious bounds on NP in $B-\Bbar$ mixing.

\item $|V_{cb}| = (41.5 \pm 0.7)\times 10^{-3}$, $m_b^{1S} = 4.68\pm
0.03\,\GeV$, $\ov m_{c}(m_c) = 1.22\pm0.06\,\GeV$ reached unprecedented
precision and robustness, as all hadronic inputs are determined from data.

\item $A_{K^- \pi^+} = -0.12 \pm 0.02$ implies that there is large direct CPV,
so "$B$-superweak" models are excluded, and there are sizable strong phases in
some $B$ decays.

\item Much more: improvements in $|V_{ub}|$; observation of $B\to
X_s\ell^+\ell^-$, $D^{0(*)}\pi^0$, new $D_s$ \& $c\bar c$ states.

\end{itemize}\vspace*{-8pt}
The next few years promise the hope of similarly interesting results (in
arbitrary order):

\begin{itemize}\vspace*{-8pt}\itemsep -2pt

\item Clarify agreement / disagreement between $S_{\eta' K}$, $S_{\phi K}$ and
$\sin2\beta$: the current central values with $5\sigma$ significance would
signal NP.

\item Improvements in determination of $\alpha$ and $\gamma$: decrease errors,
clarify\,/\,eliminate assumptions in the analyses, will significantly improve
bounds on NP.

\item Reduce error of $|V_{ub}|$ (approach current rigor of $|V_{cb}|$): the
side opposite to $\beta$, so any progress directly improves accuracy of CKM
tests (error with continuum methods asymptote to $5\%$).

\item Achieve theoretical limits in $B\to X_s\gamma$, $B\to X_s\ell^+\ell^-$:
will impact model building, continuum theory is most precise for inclusive
decays; cannot be done well at LHCB.

\item Approach SM predictions from current $A_{\rm SL} = -(3.0 \pm 7.8)\times
10^{-3}$ and $S_{K^*\gamma} = -0.13 \pm 0.32$ measurements: these are important
to constrain certain type of extensions of the SM.

\item Firmly establish $B\to \rho\gamma$ and $B\to \tau\nu$: these are not yet
seen operators.

\item Test if the $B_s$ mixing amplitude is consistent with the SM, i.e.,
whether both that $\Delta m_{B_s}$ and $S_{B_s\to\psi\phi}$ are in the SM range
(the CKM fit predicts $\sin2\beta_s = 0.0346^{+0.0026}_{-0.0020}$).

\item The unexpected ones: similar to the "new" $c\bar s$ and $c\bar c$ states
discovered by the $B$ factories, new physics could also be discovered in the
charm sector.  Nothing forbids the possibility of seeing a clear sign of NP in
$D\to \pi\ell^+\ell^-$ or $CP$ violation in $D-\Dbar$ mixing "any time".

\end{itemize}\vspace*{-8pt}

\section*{Acknowledgments}

I am grateful to Andreas H\"ocker, Heiko Lacker, Yossi Nir, Gilad Perez, Dan
Pirjol, and Iain Stewart for many interesting discussions.  Special thanks to
Stephane Monteil and Arnaud Robert for their help with CKMfitter, and for
putting up with my questions beyond any reasonable limit.
I thank the organizers for the invitation to this very enjoyable conference, the
generous hospitality, and the excellent pub guide.
This work was supported in part by the Director, Office of Science, Office of
High Energy Physics, of the U.S.\ Department of Energy under Contract
DE-AC02-05CH11231 and by a DOE Outstanding Junior Investigator award.

\end{document}